\documentclass[journal]{IEEEtran}
\usepackage{amsmath}
\usepackage{amsfonts,amssymb}
\usepackage{graphicx}
\allowdisplaybreaks[4]
\usepackage{xcolor}
\usepackage{soul}
\definecolor{deepred}{RGB}{205,38,38}
\usepackage{algorithm}
\usepackage{algorithmic}
\usepackage{amsthm}

\usepackage{lineno,hyperref}
\modulolinenumbers[5]
\usepackage{bm}

\usepackage{multirow}
\usepackage{array}
\usepackage{booktabs}
\usepackage{lipsum}
\usepackage{etoolbox}

\makeatletter
\newcount\SOUL@minus 
\makeatother

\ifCLASSINFOpdf

\else

\fi

\ifCLASSOPTIONcompsoc
  \usepackage[caption=false,font=normalsize,labelfont=sf,textfont=sf]{subfig}
\else
  \usepackage[caption=false,font=footnotesize]{subfig}
\fi

\begin{document}
\bibliographystyle{ieeetr} 

\title{An Online Feedback-Based Linearized Power Flow Model for Unbalanced Distribution Networks}
\author{\IEEEauthorblockN{Rui Cheng, \textit{Student Member}, \textit{IEEE}, Zhaoyu Wang, \textit{Senior Member}, \textit{IEEE}, Yifei Guo, \textit{Member},  \textit{IEEE}
}
\thanks{The authors are with the Department of Electrical and Computer Engineering, Iowa State University, Ames, IA 50011 USA (e-mail: wzy@iastate.edu)}
}
\maketitle

\begin{abstract}
The nonlinearity, nonconvexity and phase coupling of power flow challenge the analysis and optimization of unbalanced distribution networks. To tackle this issue, this paper proposes an online feedback-based linearized power flow model for unbalanced distribution networks with both wye-connected and delta-connected loads. The online feedback-based linearized model is grounded on the first-order Taylor expansion of the branch flow model, and updates the model parameters via online feedback by leveraging the instantaneous measurements of voltages and load consumption. Exploiting the connection structure of unbalanced radial distribution networks, we also provide a unified matrix-vector compact form of the  model. The numerical tests on the IEEE 123-bus test system validate the accuracy and superiority of the proposed model. Additionally, we apply the model to an optimal power flow problem, which further demonstrates its effectiveness.
\end{abstract}

\section*{{Nomenclature}}
\subsection{Acronyms and Parameters}
\begin{IEEEdescription}[\IEEEusemathlabelsep\IEEEsetlabelwidth{${\bm v}_{\rm min},{\bm v}_{\rm max}asdfaads$}]
\item[$\bar{\bm{A}}=\text{[}\bm{A}_0,\bm{A}^T\text{]}^T$] Incidence matrix for the unbalanced radial network;
\item[$\text{BFM}$] Bus flow model;
\item[$\text{BIM}$] Bus injection model;
\item[$bp(j)$] Bus immediately preceding bus $j$ along radial network;
\item[$\text{DistFlow}$] Distribution flow;
\item[$\text{FOT}$] First-order Taylor;
\item[$\text{FPL}$] Fixed-point linearization;
\item[$\bm{G}_{{ij}}^{p}(t),\bm{G}_{{ij}}^{q}(t),\bm{u}_{ij}^p(t)$] Time-varying parameters for the real power equation in the online model;
\item[$\bm{H}_{{ij}}^{p}(t),\bm{H}_{{ij}}^{q}(t),\bm{u}_{ij}^q(t)$] Time-varying parameters for the reactive power equation in the online model;
\item[$i^{\phi}$] Phase $\phi$ node of bus $i$ and $\phi\in\Phi_i$;
\item[$\bm{K}_i(t)$] Time-varying parameter associated with delta-connected load at bus $i$ in the online model;
\item[$\ell_j=(i,j)$] Line segment connecting buses $i$ and $j$ with $i= bp(j)$ and $j \in \mathcal{N}$;
\item[$\ell_j^{\phi}$] Phase $\phi$ circuit of $\ell_j$ and $\phi\in\Phi_{ij}$;
\item[$\text{LinDistFlow}$] Linearized distribution flow;
\item[$\bm{M}_{{ij}}^{p}(t),\bm{M}_{{ij}}^{q}(t),\bm{u}_{ij}^v(t)$] Time-varying parameters for the voltage equation in the online model;
\item[$\text{MAPE}$] Mean Absolute Percentage Error;
\item[$n_i$] Number of phases for bus $i$;
\item[$n_{\ell_j}$] Number of phases for line segment $\ell_j$;
\item[$n_{\Delta,i}^{\phi\phi'}$] Number of phase-to-phase connections in $\Phi\Phi'_i$;
\item[$\text{OPF}$] Optimal power flow;
\item[$\bm{p}_g$] Real power vector of PV generators for all buses in $\mathcal{N}$;
\item[$\bm{q}_g^{min},\bm{q}_g^{max}$] Lower and upper bounds for reactive power of PV generators;
\item[$\bm{s}_{Y,i}=\bm{p}_{Y,i}+j\bm{q}_{Y,i}$] Net complex power consumption of wye-connected load at bus $i$;
\item[$\bm{s}_Y=\bm{p}_{Y}+j\bm{q}_{Y}$] Net complex power consumption of wye-connected load for all buses in $\mathcal{N}$;
\item[$\bm{s}_{\Delta,i}^{\Phi\Phi'}=\bm{p}_{\Delta,i}^{\Phi\Phi'}+j\bm{q}_{\Delta,i}^{\Phi\Phi'}$] Phase-to-phase net complex power consumption of delta-connected load at bus $i$;
\item[$\bm{s}_{\Delta}^{\Phi\Phi'}=\bm{p}_{\Delta}^{\Phi\Phi'}+j\bm{q}_{\Delta}^{\Phi\Phi'}$] Phase-to-phase net complex power consumption of delta-connected load for all buses in $\mathcal{N}$;
\item[$\text{VVC}$] Volt-VAr control;
\item[$\bm{z}_{ij}=\bm{r}_{ij}+j\bm{x}_{ij}$] Impedance matrix for line segment $\ell_j=(i,j)$;
\end{IEEEdescription}
\subsection{Sets}
\begin{IEEEdescription}[\IEEEusemathlabelsep\IEEEsetlabelwidth{${\bm v}_{\rm min},{\bm v}_{\rm max}asdfaads$}]
\item[$\mathcal{L}$] Edge set of line segments;
\item[$\mathcal{N}$] Index set for all non-head buses of the radial network;
\item[$\mathcal{N}_j$] Index set for all buses that follow bus $j$ but exclude bus $j$;
\item[$\Phi_i\subseteq\{a,b,c\}$] Phase set of bus $i$;
\item[$\Phi_{ij}\subseteq\{a,b,c\}$] Phase set of line segment $\ell_j=(i,j)$;
\item[$\Phi\Phi'_i\subseteq\{ab,bc,ca\}$] Set of phase-to-phase connections for the delta-connected load at bus $i$;
\end{IEEEdescription}
\subsection{Variables}
\begin{IEEEdescription}[\IEEEusemathlabelsep\IEEEsetlabelwidth{${\bm v}_{\rm min},{\bm v}_{\rm max}asdfaads$}]
\item[$\bm{d}_{ij}^v,\bm{d}_{ij}^p,\bm{d}_{ij}^q$] Nonlinear voltage drop term, real and reactive power loss terms;
\item[$\text{inv}(\bm{V}_{i}^{\Phi_{ij}})$] Column vector consisting of $\frac{1}{{V}_i^{\phi}}$ for $\phi\in\Phi_{ij}$;
\item[$I_{ij}^{\phi}$] Phase $\phi$ current from bus $i$ to bus $j$ and $\phi\in\Phi_{ij}$; 
\item[$\bm{I}_{ij}$] Current vector from bus $i$ to bus $j$;
\item[$\bm{P},\bm{Q}$] Real \& Reactive power flow vectors over all line segments;
\item[$\bm{q}_g$] Reactive power vector of PV generators for all buses;
\item[$\bm{T}_{\Delta,i}$] Transformation matrix for the delta-connected load at bus $i$; 
\item[${\bm{s}_{\Delta,i}=\bm{p}_{\Delta,i}+j\bm{q}_{\Delta,i}}$] Power flow vector from bus $i$ to the delta-connected load;
\item[$\bm{s}_i=\bm{p}_i+j\bm{q}_i$] Net complex power consumption for bus $i$;
\item[$\bm{s}=\bm{p}+j\bm{q}$] Net complex power consumption vector for all buses;
\item[$\bm{S}_{ij}=\bm{P}_{ij}+j\bm{Q}_{ij}$] Power flow vector from bus $i$ to $j$;
\item[$\bm{v}_{i}$] Squared voltage magnitude vector of bus $i$;
\item[$\bm{v}_{i}^{\Phi_{ij}}$] Sub-vector of $\bm{v}_i$ consisting of entries associated with $\Phi_{ij}$;
\item[$\bm{v}$] Squared voltage magnitude vector of all buses in $\mathcal{N}$;
\item[$\bm{V}_{i}$] Complex voltage vector of bus $i$;
\item[$\bm{V}_{i}^{\Phi_{ij}}$] Sub-vector of $\bm{V}_i$ consisting of entries associated with $\Phi_{ij}$;
\item[$\bm{V}$] Complex voltage vector for all buses in $\mathcal{N}$;
\end{IEEEdescription}
\subsection{Operator}
\begin{IEEEdescription}[\IEEEusemathlabelsep\IEEEsetlabelwidth{${\bm v}_{\rm min},{\bm v}_{\rm max}asdfaaa$}]
\item[$\text{blkdiag}(\bm{U}_1,...,\bm{U}_N)$] A block diagonal matrix created by aligning the matrices $\bm{U}_1,...,\bm{U}_N$ along its diagonal. 
\item[$\text{diag}(\bm{u})$] A square matrix with the entries of $\bm{u}$ in its diagonal
\item[$\text{imag}\text{[}\bm{u}\text{]}$] Imaginary part of $\bm{u}$
\item[$\text{real}\text{[}\bm{u}\text{]}$] Real part of $\bm{u}$
\item[$\oslash$] Element-wise division
\item[$\odot$] Element-wise multiplication
\item[$|\cdot|$] Element-wise magnitude operation
\item[$|\cdot|^2$] Element-wise square operation
\item[$(\cdot)^*$] Element-wise complex-conjugate operation
\item[$(\cdot)^T$] Transposition operation
\item[$(\cdot)^H$] Complex-conjugate transposition operation
\end{IEEEdescription}

\section{Introduction}
\IEEEPARstart{P}{ower} flow modeling is the fundamental in power system analysis, optimization, and control. However, the nonlinear and nonconvex nature of power flow  poses great challenges for high-efficiency computation and optimization. In particular, power flow models in distribution networks are even more complex due to the unbalanced operation. 

To address the challenges, convex relaxation and linearization approaches have been proposed and investigated in recent years. In general, convex relaxation approaches can be classified into second-order cone program relaxations \cite{RAJ}-\cite{BFM}, semidefinite relaxations \cite{XB}-\cite{EDAHZ}, chordal relaxations \cite{MAS}-\cite{SB}. See \cite{BFM2} for a tutorial of convex relaxation methods for balanced networks. 

Compared to convex relaxation approaches, linerization approaches have lower complexity and  higher computational efficiency, which have attracted increasing attentions. DC power flow is one of the popular linerization approaches for analysis and operation of electric power systems \cite{BS}. However, DC power flow is not applicable to distribution networks since the resistive and reactive parts of the line impedance in distribution networks are comparable. In \cite{SVD}, linearization of nonlinear power flow equations, based on the bus injection model (BIM) with voltages expressed in rectangular coordinates, is studied. The classical distribution flow (\textit{DistFlow}), based on the branch flow model (BFM), is proposed in \cite{MBF}, which is regarded as a well-established method for recursively solving the power flow in single-phase distribution networks. However, the non-linear power loss term in DistFlow leads to an non-convex formulation. To solve this problem, the linearized distribution flow (\textit{LinDistFlow}) is developed by neglecting the non-linear power loss term in \cite{MB}-\cite{MB2}, which could be a good approximation when the power loss is much smaller than the branch power flow. The compact LinDistFlow representation using graph-based matrices for single-phase radial distribution networks is proposed in \cite{MF}-\cite{VK}. Nevertheless, the above works \cite{BS}-\cite{VK} can only be applied to single-phase distribution networks without taking into account the phase coupling in unbalanced distribution networks. 

For unbalanced distribution networks, linearized power flow models have been proposed in \cite{AG}-\cite{LG}. A linear power flow model for three-phase distribution systems, based on a rectangular formulation of BIM, is proposed in \cite{AG}. But this model is not suitable for applications of optimal power flow (OPF) since it includes the product of load and voltage variables, leading to the non-convex OPF formulation. The LinDistFlow model is further extended to multiphase unbalanced distribution networks in \cite{BAR}-\cite{LG},
{which has been widely used in different areas in power systems. For example, the extended LinDistFlow model is applied to solve voltage regulation problem in {\cite{XZ}}.} 
However, there are two major limitations in \cite{BAR}-\cite{LG}: (i) The extended LinDistFlow model for unbalanced distribution networks is based on a relatively strong assumption that the phase voltages across networks are nearly balanced, which is usually difficult to meet in reality; (ii) Only wye-connected loads are considered. In practice, most distribution grids are  multiphase, radial networks with both wye-connected and delta-connected loads. {A lossy LinDistFlow formulation is proposed for both single-phase and multi-phase distribution networks {\cite{ES}}, which estimates the line losses via parametrization to improve performance of the extended LinDistFlow. However, the results in {\cite{ES}} show that the lossy LinDistFlow performs much better for single-phase networks than multi-phase networks, its accuracy improvement for multi-phase distribution networks is limited compared with the extended LinDistFlow.} Moreover, all of the above linear power flow models \cite{{BS}}-\cite{ES} are essentially \textit{offline} methods. As the penetration of distributed energy resources increases, distribution power flow can change rapidly over time. However, offline models cannot capture and track the \textit{time-varying} system characteristics due to its open-loop nature, thus, potentially leading to non-negligible errors and inaccurate power flow solutions.

Recently, a great number of sensors, such as advanced metering infrastructure and micro phasor measurement units, have been deployed in distribution networks. Thanks to the significantly enhanced monitoring capability and observability, {it makes the online power flow modeling very promising, by taking advantage of measurements as feedback to establish an online updated model. Recent years have seen a dramatic surge of interest in the online power flow and its associated applications. In {\cite{LO}}-{{\cite{AH}}}, online feedback optimizations are proposed to solve different power system problems (e.g., voltage regulation problems) in single-phase networks. Approximate linear models, developed from the BIM, have been recently utilized to develop real-time OPF solvers for single-phase distribution systems {\cite{EDA}}, {\cite{EDA2}}. In {\cite{AB}}-{\cite{AB2}}, the online first-order Taylor (FOT) model and fixed-point linearization (FPL) model, based on BIM, are proposed to better adapt to the fast changes in generic unbalanced distribution networks, and they thus can be utilized to broaden the applicability of {\cite{EDA}}, {\cite{EDA2}}.} Compared to the FPL model, the FOT model can provide a better local linear approximation, but it always suffers high computational complexity because a large number of equations need to be solved to update the parameters, which hinders the applicability of the online FOT model. Unlike BIM, BFM has the advantage that its variables correspond directly to physical quantities, such as branch power, and therefore are often more intuitive than BIM \cite{BFM}. However, to date, the branch-based online model for unbalanced distribution power flow has not been well investigated.

To resolve these problems,  this paper proposes an online feedback-based linearized power flow model based on BFM for unbalanced distribution networks. The proposed model is designed to update its parameters online by leveraging the instantaneous measurements of voltages and load consumption to guarantee the accuracy. {Compared to existing methods, the main contributions of this paper are summarized as follows: (i) Unlike online power flow models derived from BIM, the proposed online model is based on BFM. It can be applicable to unbalanced distribution networks with both wye-connected and delta-connected loads, which broadens the application of BFM in online power flow.
(ii) Taking advantage of measurements, this branch-based online model can better capture the time-varying characteristics (e.g. the multi-phase imbalance) in distribution networks compared with the conventional LinDistFlow models.
(iii) This online model is essentially grounded on the FOT expansion of BFM. All the parameters of this model, represented by the closed-form analytic expressions, can be updated without solving equation sets. Hence, it has an inherently low computational complexity, which is suitable for online implementation.
(iv) The graph representation of LinDistFlow for single-phase radial distribution networks {\cite{VK}} is extended to unbalanced distribution networks. Exploiting the connection structure of unbalanced radial distribution networks, a unified matrix-vector compact form of this online model is also provided. 
}

The remainder of this paper is organized as follows. The exact nonlinear power flow formulation for unbalanced distribution networks with wye-connected and delta-connected loads is discussed in Section \ref{sec:PowerFlow}. The online feedback-based linearized power flow model, its compact form as well as potential applications are described in Section \ref{sec:DataDriven}. The effectiveness and superiority of our proposed model are verified in Section \ref{sec:Case} and concluding comments are given in Section~\ref{sec:Conclusion}.

\section{Exact Power Flow Formulation}
\label{sec:PowerFlow}

\subsection{A Standard Nonlinear Power Flow Model}
\begin{figure}[htb]
    \centering
    \includegraphics[width=3.3in]{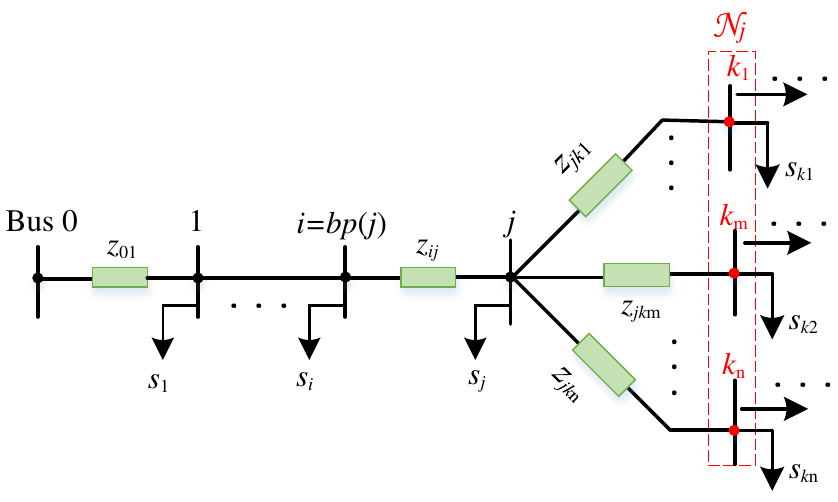}
    \caption{A radial distribution network}
    \label{fig:RadialNetwork}
\end{figure}

Consider a radial distribution network with $N$+1 buses. Let $\{0\}\bigcup\mathcal{N}$ denote the index set for these buses, where ${\mathcal{N}}:=\{1,2,...,N\}$. As depicted in Fig.\ref{fig:RadialNetwork}, for each bus $j\in\mathcal{N}$, let $bp(j)$ denote the bus that immediately precedes bus $j$ along the radial network headed by bus 0. Also, let $\mathcal{N}_j$ denote the set of all buses $n\in\mathcal{N}$ that follow bus $j$ but exclude bus $j$. Let $\mathcal{L}:=\{\ell_j=(i,j)|i=bp(j), j\in\mathcal{N}\}$ denote this edge set of line segments.  Note that there is only a unique line segment $\ell_j$ for $\forall{j}\in\mathcal{N}$ due to the radial network topology. 

For each bus $i\in\mathcal{N}\bigcup\{0\}$, let
$\Phi_i$ denote the phase set of bus $i$, $i^{\phi}$ denote the phase $\phi$ node of bus $i$ for $\phi\in\Phi_i$, $n_{i}$ denote the number of phases for bus $i$. Denote by $V_i^\phi$, $|V_i^\phi|$, $s_{i}^\phi=p_{i}^\phi+jq_{i}^\phi$ the complex voltage, the voltage magnitude, the net complex power consumption of $i^{\phi}$, respectively. And define column vectors $\bm{V}_i:=[V_{i}^{\phi}]_{\phi\in\Phi_{i}}$, $|\bm{V}_i|:=[|V_i^{\phi}|]_{\phi\in\Phi_{i}}$, $\bm{v}_i:=|\bm{V}_i|^2$, $\bm{s}_{i}:=[\bm{s}_i^\phi]_{\phi\in\Phi_{i}}$, where $\bm{s}_{i}=\bm{p}_{i}+j\bm{q}_{i}$. For each line segment $\ell_j=(i,j)\in\mathcal{L}$, let $\Phi_{ij}$ denote the phase set of line segment $\ell_j$, $\ell_j^\phi$ denote the phase $\phi$ circuit of $\ell_j$ for $\phi\in\Phi_{ij}$, $n_{\ell_j}$ denote the number of phases for  line segment $\ell_j$. Denote by $I_{ij}^{\phi}$, $S_{ij}^{\phi}=P_{ij}^{\phi}+jQ_{ij}^{\phi}$ the current and the power flow over $\ell_j^\phi$. And define column vectors 
$\bm{I}_{ij}:=[I_{ij}^\phi]_{\phi\in\Phi_{ij}}$, $\bm{S}_{ij}:=[S_{ij}^\phi]_{\phi\in\Phi_{ij}}$, where $\bm{S}_{ij}=\bm{P}_{ij}+j\bm{Q}_{ij}$, $\bm{V}_i^{\Phi_{ij}}:=[{V}_i^{\phi}]_{\phi\in\Phi_{ij}}$, $|\bm{V}_i^{\Phi_{ij}}|:=[|V_i^{\phi}|]_{\phi\in\Phi_{ij}}$,  $\bm{v}_i^{\Phi_{ij}}=|\bm{V}_i^{\Phi_{ij}}|^2$, $\text{inv}(\bm{V}_i^{\Phi_{ij}}):=[\frac{1}{{V}_i^{\phi}}]_{\phi\in\Phi_{ij}}$. And let $\bm{z}_{ij}=\bm{r}_{ij}+j\bm{x}_{ij}\in\mathbb{C}^{n_{\ell_j}\times{n}_{\ell_j}}$ denote its impedance matrix for line segment $\ell_j$.

For any $(i,j)\in\mathcal{L}$, the multi-phase BFM can be represented as follows \footnote{Note that for $\forall(i,j)\in\mathcal{L}$, $\bm{V}_i^{\Phi_{ij}}$ could be different from $\bm{V}_i$, since $\Phi_{ij}$ may not be the same as $\Phi_{i}$, but $\bm{V}_j^{\Phi_{ij}}$ is the same as $\bm{V}_j$.}:
\begin{subequations}\label{eq:V}
\begin{align}
    \bm{V}_j&=\bm{V}_i^{\Phi_{ij}}-\bm{z}_{ij}\bm{I}_{ij}\\\bm{I}_{ij}&=\bm{S}_{ij}^*\oslash(\bm{V}_i^{\Phi_{ij}})^*\\
    \bm{S}_{ij}&=\sum_{k\in\mathcal{N}_j}\bm{S}_{jk}+\bm{s}_j+\bm{I}_{ij}^*\odot(\bm{z}_{ij}\bm{I}_{ij})
\end{align}
\end{subequations}

\subsection{Wye-Connected and Delta-Connected Load Models}

For the wye-connected load, let $s_{Y,i}^\phi=p_{Y,i}^\phi+jq_{Y,i}^\phi$ denote the net complex power consumption of bus $i$ at $\phi\in\Phi_i$ and define the column vector $\bm{s}_{Y,i}:=[{s}_{Y,i}^\phi]_{\phi\in\Phi_{i}}$, where $\bm{s}_{Y,i}=\bm{p}_{Y,i}+j\bm{q}_{Y,i}$. For the delta-connected load, let $\Phi\Phi'_i\subseteq\{ab,bc,ca\}$ denote the set of phase-to-phase connections for the delta-connected load at bus $i$, $n_{\Delta,i}^{\Phi\Phi'}$ denote the number of phase-to-phase connections in $\Phi\Phi'_i$, ${s}_{\Delta,i}^{\phi\phi'}={p}_{\Delta,i}^{\phi\phi'}+j{q}_{\Delta,i}^{\phi\phi'}$ denote its phase-to-phase $\phi\phi'$ net complex power consumption and define the column vector $\bm{s}^{\Phi\Phi'}_{\Delta,i}:=[{s}_{\Delta,i}^{\phi\phi'}]_{\phi\phi'\in\Phi\Phi'_i}$, where $\bm{s}^{\Phi\Phi'}_{\Delta,i}=\bm{p}^{\Phi\Phi'}_{\Delta,i}+j\bm{q}^{\Phi\Phi'}_{\Delta,i}\in\mathbb{C}^{n_{\Delta,i}^{\Phi\Phi'}}$. Let $s_{\Delta,i}^\phi=p_{\Delta,i}^\phi+jq_{\Delta,i}^\phi$ denote the phase $\phi$ power flow from bus $i$ to the delta-connected load, and define the column vector $\bm{s}_{\Delta,i}=[s_{\Delta,i}^\phi]_{\phi\in\Phi_i}$, where $\bm{s}_{\Delta,i}=\bm{p}_{\Delta,i}+j\bm{q}_{\Delta,i}\in\mathbb{C}^{n_i}$. In general, the transition between $\bm{s}_{\Delta,i}$ and $\bm{s}^{\Phi\Phi'}_{\Delta,i}$ can be represented in the following manner: 
\begin{equation}\label{eq:DeltaTrans}
    \bm{s}_{\Delta,i}=\bm{T}_{\Delta,i}\bm{s}^{\Phi\Phi'}_{\Delta,i}, \forall{i}\in\mathcal{N}
\end{equation}
where $\bm{T}_{\Delta,i}\in\mathbb{C}^{n_i\times{n_{\Delta,i}^{\Phi\Phi'}}}$ is a transformation matrix, related to the voltage $\bm{V}_i$, for the delta-connected load at bus $i$, its dimension varies with the connection structure of delta-connected load.
\begin{figure}[htb]
    \centering
    \includegraphics[width=3in]{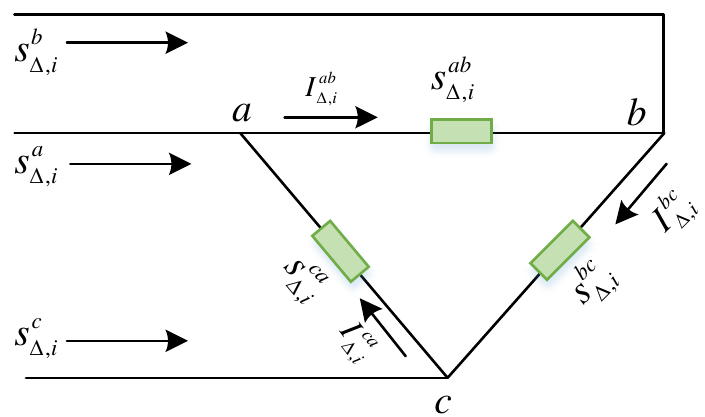}
    \caption{Structure of closed delta-connected load.}
    \label{fig:DeltaLoad}
\end{figure}


We take a closed delta-connected load as an example for illustration. As show in Fig.\ref{fig:DeltaLoad}, for the closed delta-connected load, $\bm{s}_{\Delta,i}$ and $\bm{s}^{\Phi\Phi'}_{\Delta,i}$ satisfy:


\begin{equation}
    \begin{bmatrix}
    s_{\Delta,i}^{a}\\
    s_{\Delta,i}^{b}\\
    s_{\Delta,i}^{c}
    \end{bmatrix}
    =
    \begin{bmatrix}
    \frac{V_i^a}{V_i^a-V_i^b}&0&-\frac{V_i^a}{V_i^c-V_i^a}\\
    -\frac{V_i^b}{V_i^a-V_i^b}&\frac{V_i^b}{V_i^b-V_i^c}&0\\
    0&-\frac{V_i^c}{V_i^b-V_i^c}&\frac{V_i^c}{V_i^c-V_i^a}
    \end{bmatrix}
    \begin{bmatrix}
     s_{\Delta,i}^{ab}\\
    s_{\Delta,i}^{bc}\\
    s_{\Delta,i}^{ca}
    \end{bmatrix}\label{eq:DeltaLoadMatrix}
\end{equation}
Note that (\ref{eq:DeltaLoadMatrix}) can be easily and flexibly extend to open delta-connected loads by setting the missing voltages and currents to zero. The transition between $\bm{s}_{\Delta,i}$ and $\bm{s}^{\Phi\Phi'}_{\Delta,i}$ for open delta-connected loads is shown in Appendix A.

Without loss of generality, we assume that each bus $i\in\mathcal{N}$ has both wye-connected and delta-connected loads, we have:
\begin{equation}\label{eq:DeltaYLoad}
    \bm{s}_i=\bm{s}_{Y,i}+\bm{s}_{\Delta,i}=\bm{s}_{Y,i}+\bm{T}_{\Delta,i}\bm{s}^{\Phi\Phi'}_{\Delta,i}
\end{equation}
If bus $i$ only has wye-connected loads, we set $\bm{s}_{\Delta,i}^{\Phi\Phi'}=\bm{0}$; if bus $i$ only has delta-connected loads, we set $\bm{s}_{Y,i}=\bm{0}$.

\section{An online Feedback-Based linearized Model}\label{sec:DataDriven}
Given the increasing monitoring capability and observability of distribution networks, this section proposes an online feedback-based linearized power flow model to address the non-linearity and non-convexity challenges of the unbalanced distribution power flow. To better capture the time-varying characteristics of unbalanced distribution networks, the parameters of the online feedback-based linearized model are continuously updated based on the measured voltages and load consumption at the previous time step.   
\subsection{Reformulation of Nonlinear Power Flow Model}
For the later online linearization purpose, we first reformulate (\ref{eq:V}) in this subsection. We introduce $\bm{\tilde{z}}_{ij},~\bm{\bar{z}}_{ij},~\bm{\check{z}}_{ij}\in\mathbb{C}^{n_{\ell_j}\times{n}_{\ell_j}}$ for each line segment $(i,j)\in\mathcal{L}$:
\begin{subequations}\label{eq:z}
\begin{align}
  \bm{\tilde{z}}_{ij}&=\bm{\tilde{r}}_{ij}+j\bm{\tilde{x}}_{ij}=\big[(\bm{V}_i^{\Phi_{ij}})^*\big(\text{inv}(\bm{V}_i^{\Phi_{ij}})\big)^H\big]\odot\bm{z}_{ij}\\
 \bm{\bar{z}}_{ij}&=\bm{\bar{r}}_{ij}+j\bm{\bar{x}}_{ij}=\bm{z}_{ij}\text{diag}\big((\text{inv}(\bm{V}_i^{\Phi_{ij}}))^*\big)\\
  \bm{\check{z}}_{ij}&=\bm{\check{r}}_{ij}+j\bm{\check{x}}_{ij}=\big[\text{inv}(\bm{V}_i^{\Phi_{ij}})\big(\text{inv}(\bm{V}_i^{\Phi_{ij}})\big)^H\big]\odot\bm{z}_{ij}
\end{align}
\end{subequations}
Note that $\bm{\tilde{z}}_{ij},~\bm{\bar{z}}_{ij},~\bm{\check{z}}_{ij}$ are all internally related to $\bm{V}_i^{\Phi_{ij}}$. Substituting (\ref{eq:V}b) into (\ref{eq:V}a), (\ref{eq:V}c) and taking the element-wise product of each side of (\ref{eq:V}a) with its conjugate, for $\forall(i,j)\in\mathcal{L}$, we have:
\begin{subequations}\label{bfm}
\begin{align}
    \bm{v}_{j}&=\bm{v}_i^{\Phi_{ij}}-2(\bm{\tilde{r}}_{ij}\bm{P}_{ij}+\bm{\tilde{x}}_{ij}\bm{Q}_{ij})+\bm{d}_{ij}^v\\
    &\bm{P}_{ij}=\sum_{k\in\mathcal{N}_j}\bm{P}_{jk}+\bm{p}_j+\bm{d}_{ij}^p\\
    &\bm{Q}_{ij}=\sum_{k\in\mathcal{N}_j}\bm{Q}_{jk}+\bm{q}_j+\bm{d}_{ij}^q
\end{align}
\end{subequations}
with
\begin{align*}
    \nonumber~\bm{d}_{ij}^v&=[\bm{z}_{ij}\big(\bm{S}_{ij}^*\oslash(\bm{V}_i^{\Phi_{ij}})^*\big)]\odot[\bm{z}_{ij}^*(\bm{S}_{ij}\oslash\bm{V}_i^{\Phi_{ij}})]\\
    &\nonumber=(\bm{\bar{r}}_{ij}\bm{P}_{ij})\odot(\bm{\bar{r}}_{ij}\bm{P}_{ij})+(\bm{\bar{x}}_{ij}\bm{Q}_{ij})\odot(\bm{\bar{x}}_{ij}\bm{Q}_{ij})\\&\nonumber+(\bm{\bar{x}}_{ij}\bm{P}_{ij})\odot(\bm{\bar{x}}_{ij}\bm{P}_{ij})+(\bm{\bar{r}}_{ij}\bm{Q}_{ij})\odot(\bm{\bar{r}}_{ij}\bm{Q}_{ij})\\&
    +2(\bm{\bar{r}}_{ij}\bm{P}_{ij})\odot(\bm{\bar{x}}_{ij}\bm{Q}_{ij})-2(\bm{\bar{x}}_{ij}\bm{P}_{ij})\odot(\bm{\bar{r}}_{ij}\bm{Q}_{ij})
    \\
    \nonumber \bm{d}_{ij}^p&=\bm{P}_{ij}\odot(\bm{\check{r}}_{ij}\bm{P}_{ij}+\bm{\check{x}}_{ij}\bm{Q}_{ij})\\&+\bm{Q}_{ij}\odot(\bm{\check{r}}_{ij}\bm{Q}_{ij}-\bm{\check{x}}_{ij}\bm{P}_{ij})\\
    \nonumber \bm{d}_{ij}^q&=\bm{P}_{ij}\odot(\bm{\check{x}}_{ij}\bm{P}_{ij}-\bm{\check{r}}_{ij}\bm{Q}_{ij})\\&+\bm{Q}_{ij}\odot(\bm{\check{r}}_{ij}\bm{P}_{ij}+\bm{\check{x}}_{ij}\bm{Q}_{ij})
\end{align*}
where $\bm{v}_j=|\bm{V}_j|^2,~\bm{v}_j^{\Phi_{ij}}=|\bm{V}_j^{\Phi_{ij}}|^2$, $\bm{d}_{ij}^v\in\mathbb{R}^{n_{\ell_j}}$ is the nonlinear voltage drop term, and $\bm{d}_{ij}^p$ and $\bm{d}_{ij}^q\in\mathbb{R}^{n_{\ell_j}}$ are the nonlinear real and reactive power loss terms. 

{
\textit{Remark 1:} Under the assumption that the phase voltages across the network are nearly balanced (i.e., $\frac{V_i^a}{V_i^b}\approx\frac{V_i^c}{V_i^a}\approx\frac{V_i^b}{V_i^c}\approx{e}^{j\frac{2\pi}{3}}$)
, then $\bm{\tilde{r}}_{ij}$ and $\bm{\tilde{x}}_{ij}$ are constant. Plus, the assumption that $\bm{d}_{ij}^v,\bm{d}_{ij}^p,\bm{d}_{ij}^q$ are constant (e.g., $\bm{0}$),
then 
({\ref{bfm}}) reduces to the extended LinDistFlow in {\cite{BAR}}-{\cite{LG}}.
} 
\subsection{Model Linearization via Online Feedback}
In this subsection, we propose an online feedback-based power flow model with the time-varying parameters. At time step $t$, our goal is to develop linear approximations to (\ref{eq:DeltaYLoad}) and (\ref{bfm}) based on the measured voltages and load consumption at the previous time step $t$ in the following form, for $\forall(i,j)\in\mathcal{L}$:
\begin{subequations}\label{eq:OnlineFlow}
\begin{align}
    \bm{v}_j&=\bm{v}_i^{\Phi_{ij}}+\bm{M}_{{ij}}^{p}(t)\bm{P}_{ij}+\bm{M}_{{ij}}^{q}(t)\bm{Q}_{ij}+\bm{u}_{ij}^v(t)\\
    \nonumber\bm{P}_{ij}&=\sum_{k\in\mathcal{N}_j}\bm{P}_{jk}+\bm{G}_{{ij}}^p(t)\bm{P}_{ij}\\&+\bm{G}_{{ij}}^q(t)\bm{Q}_{ij}+\bm{p}_j+\bm{u}_{ij}^p(t)\\
    \nonumber\bm{Q}_{ij}&=\sum_{k\in\mathcal{N}_j}\bm{Q}_{jk}+\bm{H}_{{ij}}^p(t)\bm{P}_{ij}\\&+\bm{H}_{{ij}}^q(t)\bm{Q}_{ij}+\bm{q}_j+\bm{u}_{ij}^q(t)\\
    \bm{s}_{i}&=\bm{s}_{Y,i}+\bm{K}_{i}(t)\bm{s}^{\Phi\Phi'}_{\Delta,i},\forall{i\in\mathcal{N}}
\end{align}
\end{subequations}

\textit{Remark 2:} In this online feedback-based linearized power flow model, $\bm{M}_{{ij}}^{p}(t)$, $\bm{M}_{{ij}}^{q}(t)$, $\bm{u}_{ij}^v(t)$, $\bm{G}_{{ij}}^p(t)$, $\bm{G}_{{ij}}^q(t)$, $\bm{u}_{ij}^p(t)$, $\bm{H}_{{ij}}^p(t)$, $\bm{H}_{{ij}}^q(t)$, $\bm{u}_{ij}^q(t)$, $\bm{K}_{i}(t)$ are the time-varying parameters of this online model, updated and calculated by the measured voltages and load consumption at the previous time step $t$.

Define the voltages, line flows, and loads as follows:
\begin{align*}
\bm{V}&=[\bm{V}_1^T,,...,\bm{V}_N^T]^T\\  \bm{P}&=[\bm{P}_{bp(1)1}^T,...,\bm{P}_{bp(N)N}^T]^T\\
\bm{Q}&=[\bm{Q}_{bp(1)1}^T,...,\bm{Q}_{bp(N)N}^T]^T\\
\bm{s}_{Y}&=\bm{p}_{Y}+j\bm{q}_{Y}=[\bm{s}_{Y,1}^T,,...,\bm{s}_{Y,N}^T]^T\\ \bm{s}_{\Delta}^{\Phi\Phi'}&=\bm{p}_{\Delta}^{\Phi\Phi'}+j\bm{q}_{\Delta}^{\Phi\Phi'}=\Big[{\bm{s}_{\Delta,1}^{\Phi\Phi'}}^T,,...,{\bm{s}_{\Delta,N}^{\Phi\Phi'}}^T\Big]^T
\end{align*}

Let $(\bm{\hat{V}}(t), \bm{\hat{s}}_{Y}(t), \bm{\hat{s}}_{\Delta}^{\Phi\Phi'}(t))$ denote a measured operating point satisfying the exact nonlinear distribution power flow (\ref{eq:V}) and (\ref{eq:DeltaYLoad}) at time step $t$. According to (\ref{eq:V}), the corresponding branch power flow $\bm{\hat{P}}_{ij}(t), \bm{\hat{Q}}_{ij}(t)$ for each line segment $(i,j)\in\mathcal{L}$ in $\bm{\hat{P}}(t), \bm{\hat{Q}}(t)$ can be calculated by $\bm{\hat{V}}(t)$:
\begin{subequations}\small
\begin{align}
    \bm{\hat{P}}_{ij}(t)&=\text{real}\Big[
    \Big(
    \bm{z}_{ij}^{-1}(\bm{\hat{V}}_i^{\Phi_{ij}}(t)-\bm{\hat{V}}_j(t))
    \Big)^*\odot\bm{\hat{V}}_i^{\Phi_{ij}}(t)
    \Big]\\
    \bm{\hat{Q}}_{ij}(t)&=\text{imag}\Big[
    \Big(
    \bm{z}_{ij}^{-1}(\bm{\hat{V}}_i^{\Phi_{ij}}(t)-\bm{\hat{V}}_j(t))
    \Big)^*\odot\bm{\hat{V}}_i^{\Phi_{ij}}(t)
    \Big]
\end{align}
\end{subequations}
And let the {squared} voltage magnitudes and the net complex power consumption are compactly denoted by:
\begin{align*}
    \bm{v}&=[\bm{v}_1^T,...,\bm{v}_N^T]^T\\
    \bm{s}&=\bm{p}+j\bm{q}=[\bm{s}_1^T,...,\bm{s}_N^T]^T
\end{align*}
Let $\bm{\hat{v}}(t)$ denote the squared voltage magnitudes corresponding to $\bm{\hat{V}}(t)$, $\bm{\hat{s}}(t)=\bm{\hat{p}}(t)+j\bm{\hat{q}}(t)$ denote the net complex power consumption corresponding to $\bm{\hat{s}}_{Y}(t),~\bm{\hat{s}}_{\Delta}^{\Phi\Phi'}(t)$.

To obtain (\ref{eq:OnlineFlow}a)-(\ref{eq:OnlineFlow}c),  we take the partial derivatives of (\ref{bfm}) with respect to $\bm{{P}}_{ij}$ and $\bm{{Q}}_{ij}$. For (\ref{bfm}a), we have:
\begin{subequations}\label{eq:v}
\begin{align}
    &\frac{\partial (\bm{v}_j-\bm{v}_i^{\Phi_{ij}})}{\partial \bm{P}_{ij}}=-2\bm{\tilde{r}}_{ij}+\bm{f}_{ij}^p\\
    \nonumber \bm{f}_{ij}^p&=\frac{\partial \bm{d}_{ij}^v}{\partial \bm{P}_{ij}}=2\bm{\bar{r}}_{ij}\text{diag}(\bm{\bar{r}}_{ij}\bm{P}_{ij})+2\bm{\bar{x}}_{ij}\text{diag}(\bm{\bar{x}}_{ij}\bm{P}_{ij})\\
    &+2\bm{\bar{r}}_{ij}\text{diag}(\bm{\bar{x}}_{ij}\bm{Q}_{ij})-2\bm{\bar{x}}_{ij}\text{diag}(\bm{\bar{r}}_{ij}\bm{Q}_{ij})\\
    &\frac{\partial (\bm{v}_j-\bm{v}_i^{\Phi_{ij}})}{\partial \bm{Q}_{ij}}=-2\bm{\tilde{x}}_{ij}+f_{ij}^q\\
    \nonumber \bm{f}_{ij}^q&=\frac{\partial \bm{d}_{ij}^v}{\partial \bm{Q}_{ij}}=2\bm{\bar{x}}_{ij}\text{diag}(\bm{\bar{x}}_{ij}\bm{Q}_{ij})+2\bm{\bar{r}}_{ij}\text{diag}(\bm{\bar{r}}_{ij}\bm{Q}_{ij})\\
    &+2\bm{\bar{x}}_{ij}\text{diag}(\bm{\bar{r}}_{ij}\bm{P}_{ij})-2\bm{\bar{r}}_{ij}\text{diag}(\bm{\bar{x}}_{ij}\bm{P}_{ij})
\end{align}
\end{subequations}
For (\ref{bfm}b), we have:
\begin{subequations}\label{eq:p}
\begin{align}
    \nonumber \bm{g}_{ij}^p&=\frac{\partial \bm{d}_{ij}^p}{\partial \bm{P}_{ij}}=\text{diag}(\bm{P}_{ij})\bm{\check{r}}_{ij}+\text{diag}(\bm{\check{r}}_{ij}\bm{P}_{ij})\\
    &+\text{diag}(\bm{\check{x}}_{ij}\bm{Q}_{ij})-\text{diag}(\bm{Q}_{ij})\bm{\check{x}}_{ij}\\
    \nonumber \bm{g}_{ij}^q&=\frac{\partial \bm{d}_{ij}^p}{\partial \bm{Q}_{ij}}=\text{diag}(\bm{Q}_{ij})\bm{\check{r}}_{ij}+\text{diag}(\bm{\check{r}}_{ij}\bm{Q}_{ij})\\
    &-\text{diag}(\bm{\check{x}}_{ij}\bm{P}_{ij})+\text{diag}(\bm{P}_{ij})\bm{\check{x}}_{ij}
\end{align}
\end{subequations}
For (\ref{bfm}c), we have:
\begin{subequations}\label{eq:q}
\begin{align}
    \nonumber \bm{h}_{ij}^p&=\frac{\partial \bm{d}_{ij}^q}{\partial \bm{P}_{ij}}=\text{diag}(\bm{P}_{ij})\bm{\check{x}}_{ij}+\text{diag}(\bm{\check{x}}_{ij}\bm{P}_{ij})
    \\&-\text{diag}(\bm{\check{r}}_{ij}\bm{Q}_{ij})+\text{diag}(\bm{Q}_{ij})\bm{\check{r}}_{ij}
    \\\nonumber \bm{h}_{ij}^q&=\frac{\partial \bm{d}_{ij}^q}{\partial \bm{Q}_{ij}}=\text{diag}(\bm{Q}_{ij})\bm{\check{x}}_{ij}+\text{diag}(\bm{\check{x}}_{ij}\bm{Q}_{ij})
    \\&+\text{diag}(\bm{\check{r}}_{ij}\bm{P}_{ij})-\text{diag}(\bm{P}_{ij})\bm{\check{r}}_{ij}
\end{align}
\end{subequations}
Recall from (\ref{eq:z}) and (\ref{eq:v})-(\ref{eq:q}) that $\bm{\tilde{r}}_{ij}$, $\bm{\tilde{x}}_{ij}$, $\bm{f}_{ij}^p$, $\bm{f}_{ij}^q$, $\bm{g}_{ij}^p$, $\bm{g}_{ij}^q$, $\bm{h}_{ij}^p$, $\bm{h}_{ij}^q$ depend on  $\bm{V}_i^{\Phi_{ij}},~\bm{P}_{ij},~\bm{Q}_{ij}$. Let $\bm{\tilde{r}}_{ij}(t)$, $\bm{\tilde{x}}_{ij}(t)$, $\bm{f}_{ij}^{p}(t)$, $\bm{f}_{ij}^{q}(t)$, $\bm{g}_{ij}^{p}(t)$, $\bm{g}_{ij}^{q}(t)$, $\bm{h}_{ij}^{p}(t)$, $\bm{h}_{ij}^{q}(t)$ denote their values corresponding to the measured operating point, where $\bm{V}_i^{\Phi_{ij}}=\bm{\hat{V}}_i^{\Phi_{ij}}(t),~\bm{P}_{ij}=\bm{\hat{P}}_{ij}(t),~\bm{Q}_{ij}=\bm{\hat{Q}}_{ij}(t)$. According to the FOT expansion of (\ref{bfm}) around the measured operating point, we obtain:
\begin{subequations}\label{eq:UpdateCoefficient}
\begin{align}
    \bm{M}_{ij}^p(t):&=-2\bm{\tilde{r}}_{ij}(t)+\bm{f}_{ij}^{p}(t)\\
    \bm{M}_{ij}^q(t):&=-2\bm{\tilde{x}}_{ij}(t)+\bm{f}_{ij}^{q}(t)\\
    \nonumber\bm{u}_{ij}^v(t):&=\bm{\hat{v}}_j(t)-\bm{\hat{v}}_i^{\Phi_{ij}}(t)\\
    &-\bm{M}_{{ij}}^{p}(t)\bm{\hat{P}}_{ij}(t)-\bm{M}_{{ij}}^{q}(t)\bm{\hat{Q}}_{ij}(t)\\
    \bm{G}_{ij}^p(t):&=\bm{g}_{ij}^{p}(t)\\
    \bm{G}_{ij}^q(t):&=\bm{g}_{ij}^{q}(t)\\
    \nonumber\bm{u}_{ij}^p(t):&=\bm{\hat{P}}_{ij}(t)-\sum_{k\in\mathcal{N}_j}\bm{\hat{P}}_{jk}(t)-\bm{\hat{p}}_j(t)\\
    &-\bm{G}_{ij}^p(t)\bm{\hat{P}}_{ij}(t)-\bm{G}_{ij}^{q}\bm{\hat{Q}}_{ij}(t)\\
    \bm{H}_{ij}^p(t):&=\bm{h}_{ij}^{p}(t)\\
    \bm{H}_{ij}^q(t):&=\bm{h}_{ij}^{q}(t)\\
    \nonumber\bm{u}_{ij}^q(t):&=\bm{\hat{Q}}_{ij}(t)-\sum_{k\in\mathcal{N}_j}\bm{\hat{Q}}_{jk}(t)-\bm{\hat{q}}_j(t)\\
    &-\bm{H}_{ij}^p(t)\bm{\hat{P}}_{ij}(t)-\bm{H}_{ij}^{q}(t)\bm{\hat{Q}}_{ij}(t)
\end{align}
\end{subequations}

Recall from (\ref{eq:DeltaLoadMatrix}) that 
$\bm{T}_{\Delta,i}$ depends on $\bm{V}_i$. Let $\bm{T}_{\Delta,i}(t)$ denote the operating transformation matrix of delta-connected load corresponding to the measured operating point, where $\bm{V}_i=\bm{\hat{V}}_i(t)$. And $\bm{K}_i(t)$ in (\ref{eq:OnlineFlow}d) can be approximately represented by:
\begin{equation}\label{eq:UpdateLoad}
    \bm{K}_i(t)=\bm{T}_{\Delta,i}(t)
\end{equation}
As seen in (\ref{eq:UpdateCoefficient})--(\ref{eq:UpdateLoad}), they provide an intuitive way to update the parameters of the online feedback-based linearized power flow model (\ref{eq:OnlineFlow}). 

\textit{Remark 3}: Note that the transformation matrix $\bm{T}_{\Delta,i}$ can be applied to extend the LinDistFlow model \cite{BAR}-\cite{LG}, only considering wye-connected loads, to unbalanced distribution networks with both wye-connected and delta-connected loads. In this case, all the entries in $\bm{T}_{\Delta,i}$ are fixed and irrelevant to $t$ since it is assumed in the LinDistFlow model that voltages are nearly balanced, i.e., $\frac{V_i^a}{V_i^b}\approx\frac{V_i^b}{V_i^c}\approx\frac{V_i^c}{V_i^a}\approx{e}^{j\frac{2\pi}{3}}$. For example, the constant transformation matrix $\bm{T}_{\Delta,i}$ for closed delta-connected loads in the extended LinDisFlow model can be represented as:

\begin{equation}
    \bm{T}_{\Delta,i}
    =\frac{\sqrt{3}}{3}
    \begin{bmatrix}
    {e}^{-j\frac{\pi}{6}}&0&{e}^{j\frac{\pi}{6}}\\
    {e}^{j\frac{\pi}{6}}&{e}^{-j\frac{\pi}{6}}&0\\
    0&{e}^{j\frac{\pi}{6}}&{e}^{-j\frac{\pi}{6}}
    \end{bmatrix}.
\end{equation}

\subsection{Compact Form}

To better formulate the online feedback-based linearized power flow model, exploiting the connection structure of unbalanced radial distribution networks, we propose a new matrix-vector compact form of this power flow model. Note that for each line segment $\ell_j=(i,j)\in\mathcal{L}$, we have $n_j=n_{\ell_j}$, indicating ${\sum_{i=1}^N{n}_i=\sum_{j=1}^Nn_{\ell_j}}$. We set $m=\sum_{i=1}^N{n}_i=\sum_{j=1}^Nn_{\ell_j}$.
And let $\bm{\bar{A}}=[\bm{A}_0, \bm{A}^T]^T$ be the incidence matrix for the unbalanced radial distribution network, where $\bm{A}_{0}^T\in\mathbb{R}^{n_0\times{m}}$ represents the connection  structure between bus 0 and each of the line segments in $\mathcal{L}$,  $\bm{A}\in\mathbb{R}^{m\times{m}}$ represents the connection  structure  between the remaining buses and each of the line segments in $\mathcal{L}$.

The row of $\bm{\bar{A}}$ corresponds to the phase $\phi$ node $i^{\phi}$ of bus $i\in\{0\}\bigcup\mathcal{N}$ for $\phi\in\Phi_i$, the column of $\bm{\bar{A}}$ corresponds to the phase $\phi$ circuit $\ell_j^{\phi}$ of line segment $\ell_j=(i,j)\in\mathcal{L}$ for $\phi\in\Phi_{ij}$. More precisely, the incidence matrix $\bm{\bar{A}}$ with an entry 1 for each “from”  phase node and -1 for each “to” phase node  corresponding to each phase circuit of line segments takes the following form:
\begin{subequations}\small
\begin{align}
    \bm{\bar{A}}&=
    \begin{bmatrix}
    \bm{J}(0,\ell_{1})&\bm{J}(0,\ell_{2})&...&\bm{J}(0,\ell_{N})\\
    \bm{J}(1,\ell_{1})&\bm{J}(1,\ell_{2})&...&\bm{J}(1,\ell_{N})\\
    \vdots&\vdots&\ddots&\vdots\\
     \bm{J}(N,\ell_{1})&\bm{J}(N,\ell_{2})&...&\bm{J}(N,\ell_{N})
    \end{bmatrix}\\
    \bm{{A}_0}^T&=
    \begin{bmatrix}
    \bm{J}(0,\ell_{1})&\bm{J}(0,\ell_{2})&...&\bm{J}(0,\ell_{N})
    \end{bmatrix}\\
    \bm{{A}}&=
    \begin{bmatrix}
    \bm{J}(1,\ell_{1})&\bm{J}(1,\ell_{2})&...&\bm{J}(1,\ell_{N})\\
    \vdots&\vdots&\ddots&\vdots\\
     \bm{J}(N,\ell_{1})&\bm{J}(N,\ell_{2})&...&\bm{J}(N,\ell_{N})
    \end{bmatrix}
\end{align}
\end{subequations}
where $\bm{J}(i,\ell_{j})\in\mathbb{R}^{n_i\times{n}_{\ell_j}}$ indicates the connection structure between bus $i$ and line segment $\ell_j$. Specifically, if the phase node $i^{{\phi}}$ ($\phi\in\Phi_i$) is in the head of the phase circuit $\ell_j^{\phi}$ ($\phi\in\Phi_{ij}$), the corresponding entry in $\bm{J}(i,\ell_{j})$ will be 1; if the phase node $i^{{\phi}}$ is in the end of the phase circuit $\ell_j^{\phi}$, the corresponding entry in $\bm{J}(i,\ell_{j})$ will be -1; otherwise, the corresponding entry will be 0. A numerical example illustrating the construction of $\bm{\bar{A}}$ for an unbalanced radial network is given in Appendix B. 

\medskip
\textit{Proposition:} The square matrix $\bm{A}$ is invertible (See Appendix C for the proof of proposition).

 \begin{algorithm}[t]
\renewcommand\baselinestretch{1}\selectfont
\small
\caption{Update the online feedback-based linearized power flow model}
\begin{algorithmic}\label{DB}
\STATE\hspace{-3mm} {\bf Initialization:} Set the time resolution $\Delta{t}$ for the model update, and the initial time step $t$.
\STATE\hspace{-1mm} Update the online feedback-based linearized power flow model by the following steps:
\STATE\hspace{-1mm} {\bf S1:} Measure $(\bm{\hat{V}}(t), \bm{\hat{s}}_{Y}(t), \bm{\hat{s}}_{\Delta}^{\Phi\Phi'}(t))$ for time step $t$.
\STATE\hspace{-1mm} {\bf S2:} Update parameters $\bm{M}_{{ij}}^{p}(t)$, $\bm{M}_{{ij}}^{q}(t)$, $\bm{u}_{ij}^v(t)$, $\bm{G}_{{ij}}^p(t)$, $\bm{G}_{{ij}}^q(t)$, $\bm{u}_{ij}^p(t)$, $\bm{H}_{{ij}}^p(t)$, $\bm{H}_{{ij}}^q(t)$, $\bm{u}_{ij}^q(t)$ using (\ref{eq:UpdateCoefficient}), $\bm{K}_{i}(t)$ using (\ref{eq:UpdateLoad}).
\STATE\hspace{-1mm} {\bf S3:} Update $\bm{M}^p(t)$, $\bm{M}^{q}(t)$, $\bm{u}^v(t)$, $\bm{G}^p(t)$, $\bm{G}^q(t)$, $\bm{u}^p(t)$, $\bm{H}^p(t)$, $\bm{H}^q(t)$, $\bm{u}^q(t)$ using (\ref{eq:CompactParametersUpdate}) to form the compact form of this online model.

\STATE\hspace{-1mm} {\bf S4:} Perform the power flow calculation for time step $t+\Delta{t}$ using (\ref{eq:OnlineUpdatePower}).
\STATE\hspace{-1mm} {\bf S5:} $t\leftarrow t+\Delta{t}$, go to $\bf S1$.
\end{algorithmic}
\end{algorithm}
\begin{figure}[t]
    \centering
    \includegraphics[width=3.4in,height=1.5in]{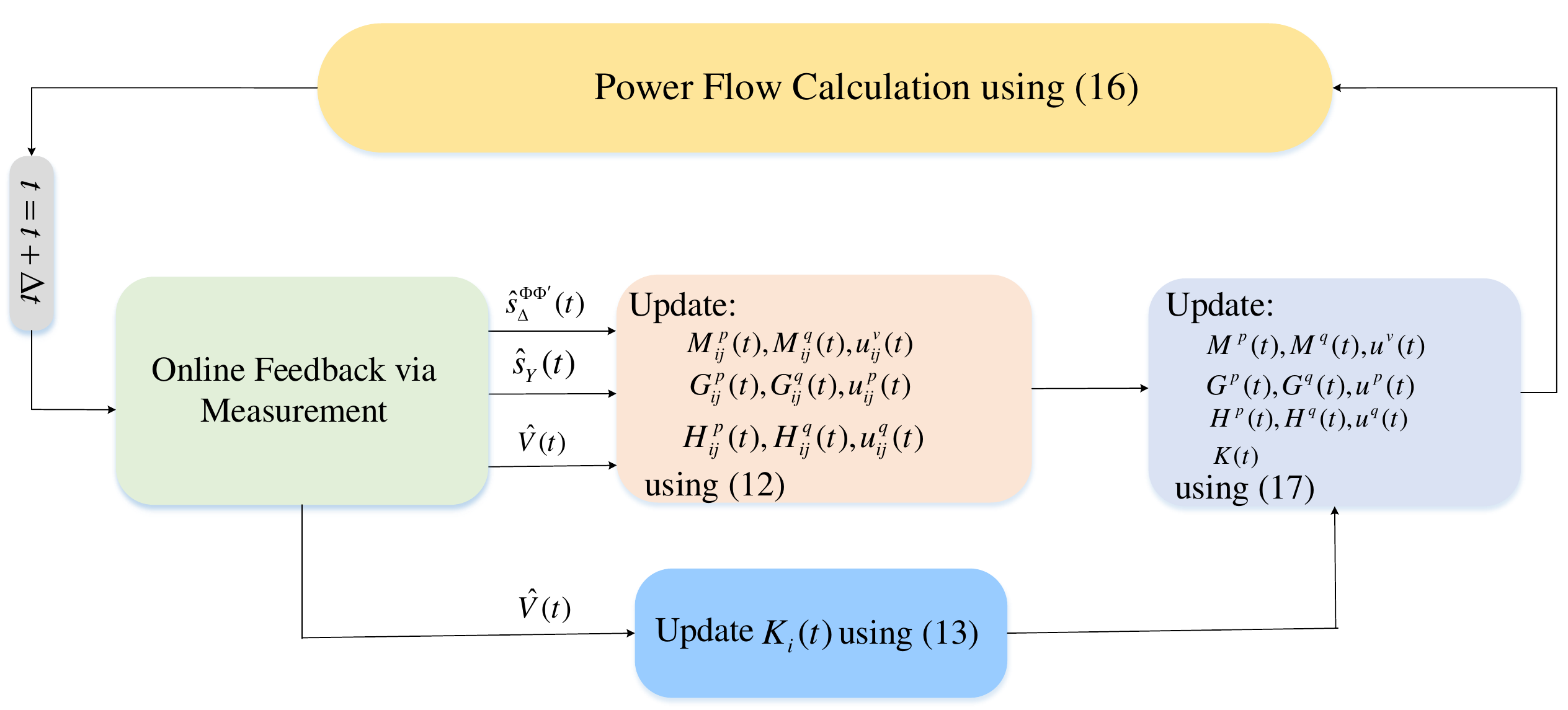}
    \caption{Model update via online feedback}
    \label{fig:OnlineFeedback}
\end{figure}

After introducing $\bm{\bar{A}}$, we can express (\ref{eq:OnlineFlow}) in a compact form as follows:
\begin{subequations}\label{eq:OnlineUpdatePower}
\begin{align}
-\begin{bmatrix}{\bm{A}}_0\,\,{\bm{A}}^T\end{bmatrix}\begin{bmatrix}{\bm{v}}_0\\{\bm{v}}\end{bmatrix}
&=[\bm{M}^p(t)\,\,\bm{M}^q(t)]
\begin{bmatrix}
\bm{P}\\
\bm{Q}
\end{bmatrix}
+\bm{u}^v(t)\\
\nonumber&\begin{bmatrix}
&\bm{A}+\bm{G}^p(t)\,\,&\bm{G}^q(t)\\
&\bm{H}^p(t)\,\,&\bm{A}+\bm{H}^q(t)
\end{bmatrix}
\begin{bmatrix}
\bm{P}\\
\bm{Q}
\end{bmatrix}
\\&+\begin{bmatrix}
\bm{u}^p(t)\\
\bm{u}^q(t)
\end{bmatrix}
=\begin{bmatrix}
-\bm{p}\\
-\bm{q}
\end{bmatrix}\\
&\bm{s}=\bm{s}_{Y}+\bm{K}(t)\bm{s}_{\Delta}^{\Phi\Phi'}
\end{align}
\end{subequations}
with
\begin{subequations}\label{eq:CompactParametersUpdate}
\begin{align}
    \bm{M}^p(t)&=\text{blkdiag}\big(\bm{M}_{{bp(1)1}}^p(t),...,\bm{M}_{{bp(N)N}}^p(t)\big)\\
    \bm{M}^q(t)&=\text{blkdiag}\big(\bm{M}_{{bp(1)1}}^q(t),...,\bm{M}_{{bp(N)N}}^q(t)\big)\\
    \bm{G}^p(t)&=\text{blkdiag}\big(\bm{G}_{{bp(1)1}}^p(t),...,\bm{G}_{{bp(N)N}}^p(t)\big)\\
    \bm{G}^q(t)&=\text{blkdiag}\big(\bm{G}_{{bp(1)1}}^q(t),...,\bm{G}_{{bp(N)N}}^q(t)\big)\\
    \bm{H}^p(t)&=\text{blkdiag}\big(\bm{H}_{{bp(1)1}}^p(t),...,\bm{H}_{{bp(N)N}}^p(t)\big)\\
    \bm{H}^q(t)&=\text{blkdiag}\big(\bm{H}_{{bp(1)1}}^q(t),...,\bm{H}_{{bp(N)N}}^q(t)\big)\\
    \bm{K}(t)&=\text{blkdiag}\big(\bm{K}_{{1}}(t),...,\bm{K}_{{N}}(t)\big)\\
    \bm{u}^v(t)&=[u_{bp(1)1}^v(t),...,u_{bp(N)N}^v(t)]^T\\
    \bm{u}^p(t)&=[u_{bp(1)1}^p(t),...,u_{bp(N)N}^p(t)]^T\\
    \bm{u}^q(t)&=[u_{bp(1)1}^q(t),...,u_{bp(N)N}^q(t)]^T.
\end{align}
\end{subequations}


As shown in Algorithm~1 and Fig.\ref{fig:OnlineFeedback}, the parameters of online linearized power flow model are updated according to the measured operating point $(\bm{\hat{V}}(t), \bm{\hat{s}}_{Y}(t), \bm{\hat{s}}_{\Delta}^{\Phi\Phi'}(t))$ at time step $t$. {The online model can be leveraged to perform the power flow and  extended to potential applications (e.g., OPF problems).}

\subsection{Potential Applications}
As the penetration of distribution energy resources increases, the traditional offline OPF that works on a slow timescale is inadequate, and we need the online OPF that can respond quickly to network changes. This proposed online feedback-based linearized power flow model (\ref{eq:OnlineUpdatePower}) can be easily integrated into the OPF related problems, e.g., Volt-VAr control (VVC), to convexify its formulation, thus facilitating the development of online OPF control algorithms in unbalanced distribution networks with both wye-connected and delta-connected loads. A simple VVC application using this online model is provided in Section \ref{sec:Case}.E.

\section{Case Study}\label{sec:Case}
Test cases are conducted in the IEEE 123 test feeder \cite{testfeeder}, an unbalanced radial distribution network, to demonstrate and verify the effectiveness and superiority of the online feedback-based linearized power flow model. The simulation results are performed with MATLAB  R2019b and the open-source Open Distribution System simulator (OpenDSS) \cite{OpenDSS}. 
               \begin{figure}[t]
   \centering
   \includegraphics[width=3.1in]{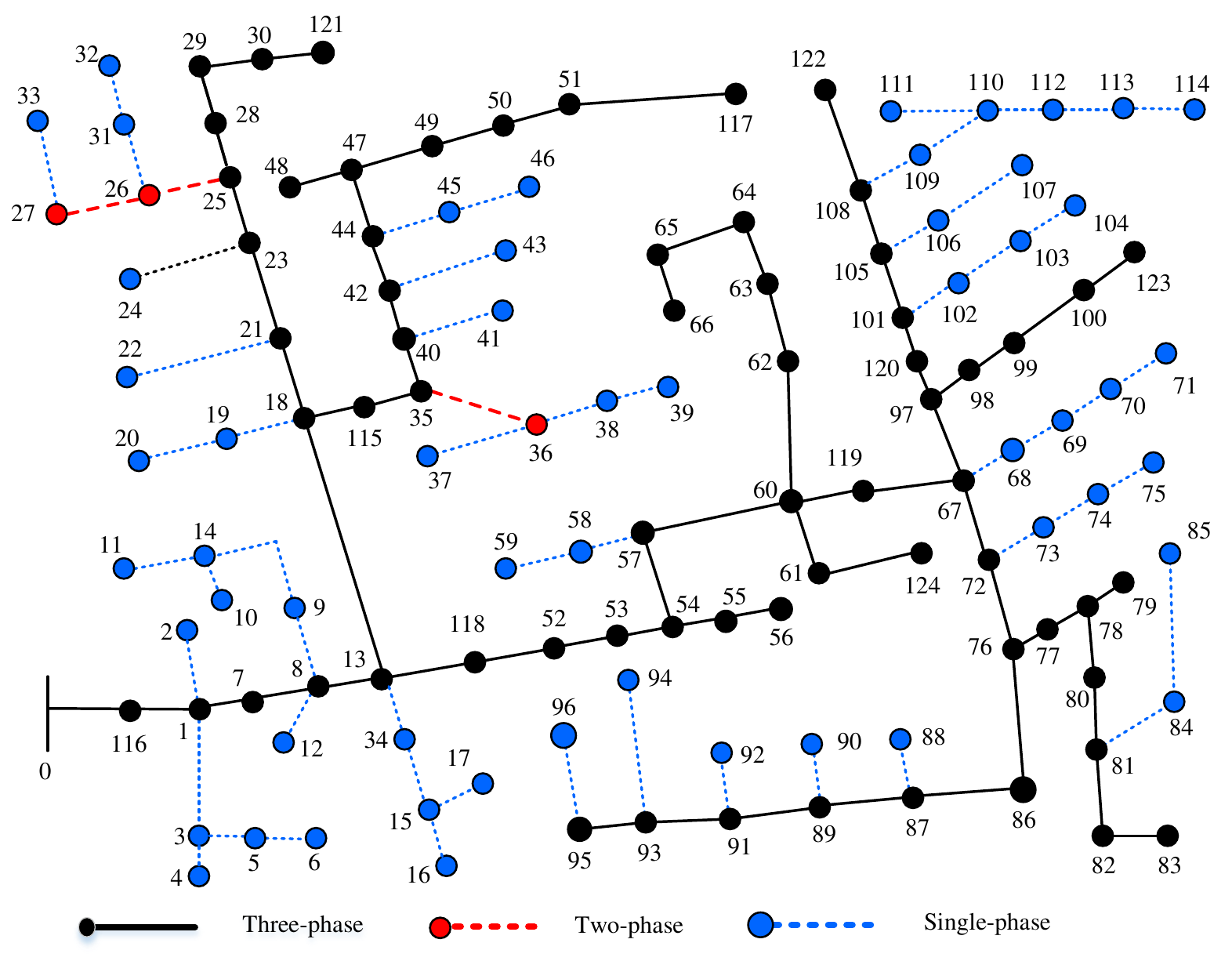}
   \caption{{IEEE 123 Bus Test Feeder}}\label{fig:IEEE123}
                        \end{figure}
\subsection{Simulation Setup}
{We first conduct a time-series simulation for the unbalanced distribution network only with wye-connected loads across one day in Section {\ref{sec:Case}}.B and {\ref{sec:Case}}.C, where the simulation time span is 24h and the time resolution of load data is set as 1min.} {The delta-connected loads are considered and discussed later in Section {\ref{sec:Case}}.D}.
The base voltage and power are set as 4.16 kV and 100 kVA, respectively. {The aggregate phase load profile across one day is shown in Fig.{\ref{fig:aggregated_load}}.
As seen in Fig.{\ref{fig:aggregated_load}}, the aggregate load of phase a is apparently greater than phase b and c. The aggregate peak load of phase a is nearly double the aggregate peak loads of phase b and c. The unbalanced load distributions among phases lead to the three-phase imbalance in the distribution network.} Consistent with the time resolution of the load data, we identify each update of the online feedback-based linearized power flow model with $\Delta{t}=1\text{min}$. \footnote{The update frequency for this online model can be set as any reasonable value in accordance with specific needs.} To verify the effectiveness and superiority of the proposed online feedback-based linearized power flow model, we compare four different methods:
\begin{figure}[t]
   \centering
   \includegraphics[width=3.2in]{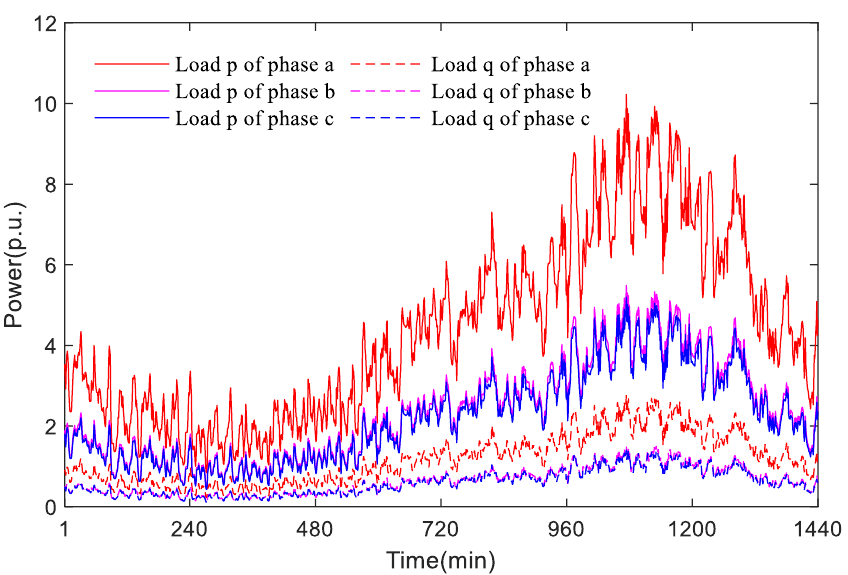}
   \caption{{Aggregate phase load across one day}}\label{fig:aggregated_load}
\end{figure}
\begin{figure}[t]
    \centering
    \includegraphics[width=3.3in,height=3in]{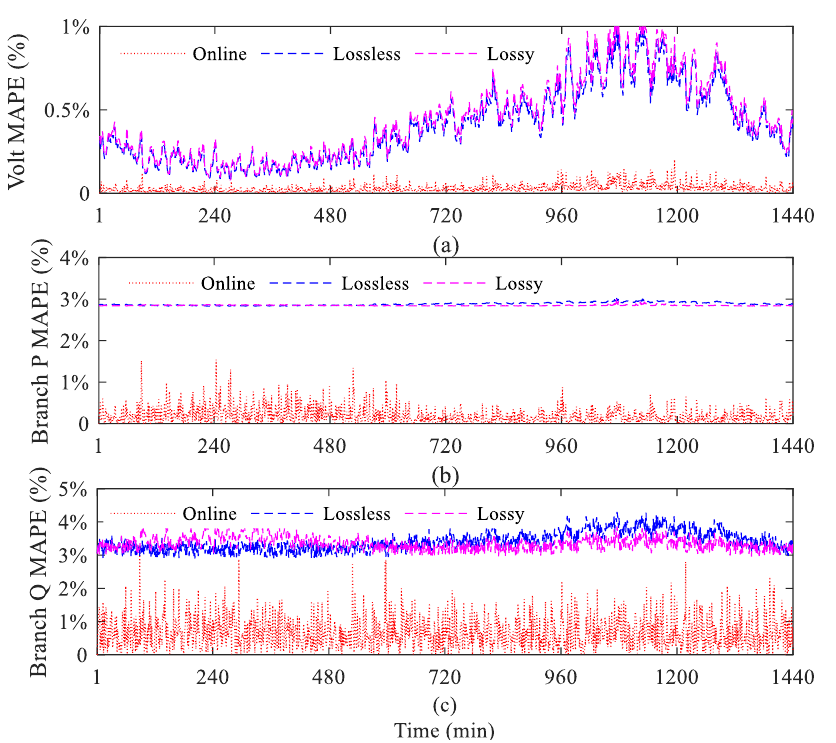}
    \caption{{MAPE values of voltage magnitude and branch power flow: (a)~Voltage magnitude MAPE; (b)~Branch real power flow MAPE; (c)~Branch reactive power flow MAPE}}
    \label{fig:error}
\end{figure}

(1) The benchmark model: The power flow solutions are calculated by OpenDSS, containing the exact nonlinear distribution power flow. 

(2) The online linearized model: The power flow solutions are calculated by the proposed online feedback-based linearized power flow model. The parameters of this online model for time step $t+\Delta{t}$ are updated by the measured voltages and load consumption at time step $t$ via online feedback.

{(3) The lossless LinDistFlow model: The power flow solutions are calculated by the extended lossless LinDistFlow proposed in {\cite{LG}}, which makes the assumption that $\frac{V_i^a}{V_i^b}\approx\frac{V_i^b}{V_i^c}\approx\frac{V_i^c}{V_i^a}\approx{e}^{j\frac{2\pi}{3}}$ and the
power loss on each radial network line segment is negligible relative to the power flow on this line segment.}

{(4) The lossy LinDistFlow model: The power flow solutions are calculated by the lossy LinDistFlow proposed in {\cite{ES}} which makes the assumption that $\frac{V_i^a}{V_i^b}\approx\frac{V_i^b}{V_i^c}\approx\frac{V_i^c}{V_i^a}\approx{e}^{j\frac{2\pi}{3}}$, but this model considers the line losses via parameterization.}
\subsection{Method Comparison}
\begin{figure}[t]
    \centering
    \includegraphics[width=3.1in]{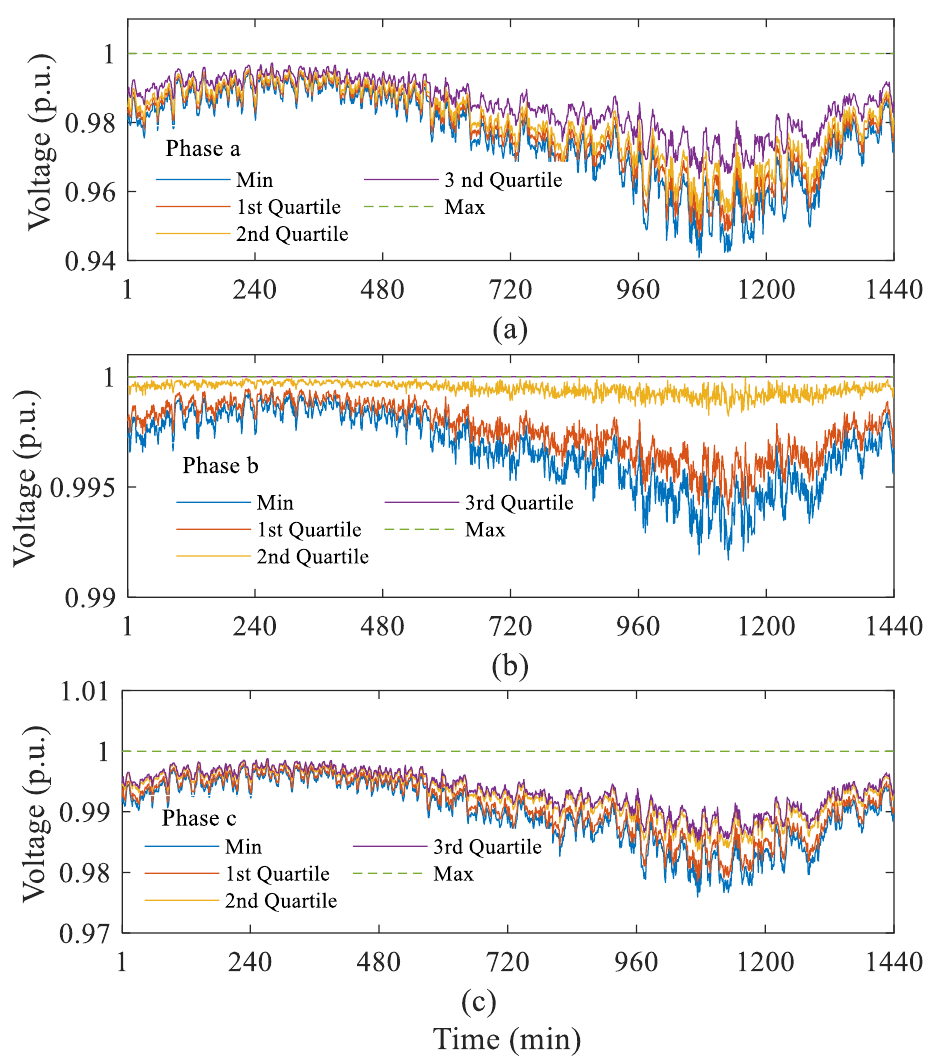}
    \caption{{Voltage magnitudes (including maximum, minimum and quartile values) of phase a, b and c at each time step : (a)~Phase a; (b)~Phase b; (c)~Phase c}}
    \label{fig:V}
\end{figure}
\begin{figure}[t]
    \centering
    \includegraphics[width=3.1in]{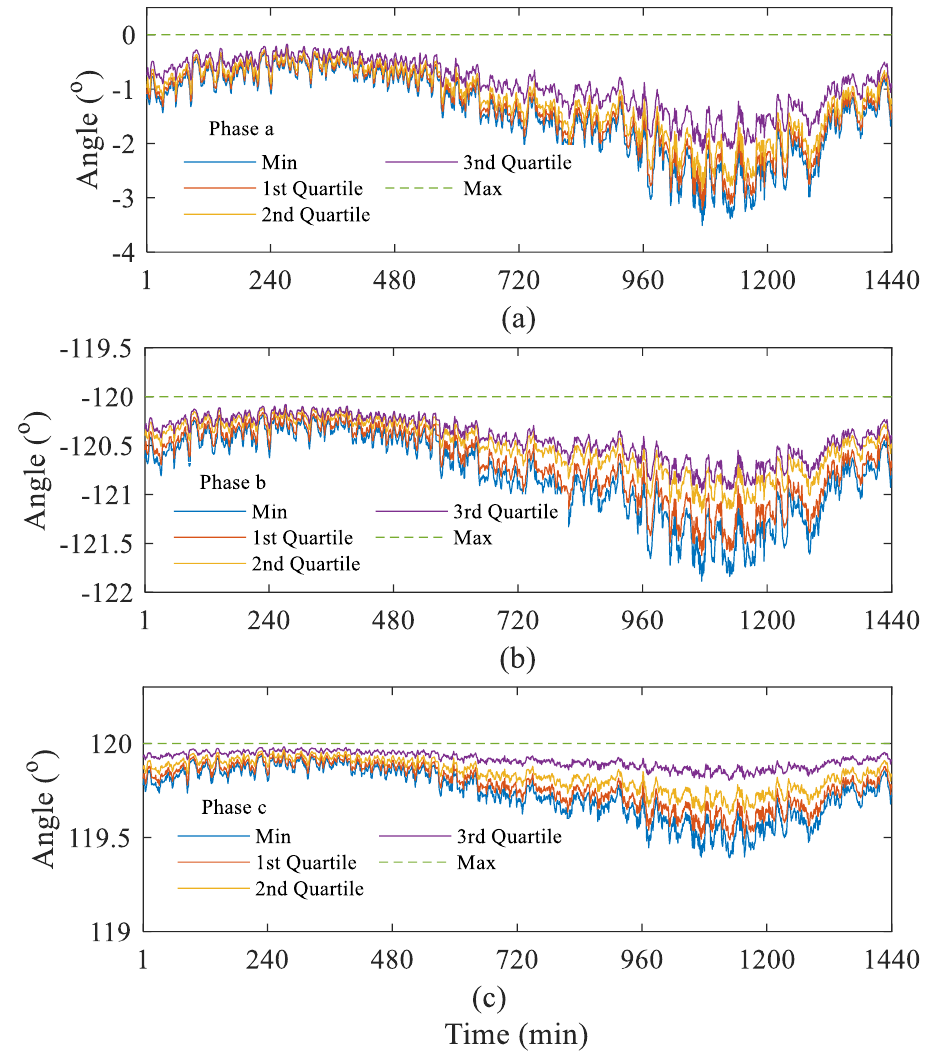}
    \caption{{Angles (including maximum, minimum and quartile values) of phase a, b and c at each time step: (a)~Phase a; (b)~Phase b; (c)~Phase c }}
    \label{fig:angle}
\end{figure}

To evaluate the accuracy of online, lossless and lossy models compared to the benchmark, the Mean Absolute Percentage Error (MAPE) is chosen as the criterion, where the benchmark values are regarded as the exact actual values and the solutions calculated by the online, lossless and lossy models are regarded as the estimated values. Fig.\ref{fig:error} shows the MAPE values of voltage magnitude and branch power flow for the online, lossless and lossy models. As shown in Fig.\ref{fig:error}, for any time step $t$, {the MAPE values of voltage magnitude and branch power flow for the online linearized model are far less than both lossless and lossy models. Maximum, minimum and quartile values of voltage magnitude and angle for phase a, b, and c at each time step $t$ are shown in Fig.{\ref{fig:V}} and Fig.{\ref{fig:angle}}.}  From Fig.\ref{fig:V} and Fig.\ref{fig:angle}, we can find the overall voltage magnitudes in phase a are clearly lower than phase b and c. In addition, the angle differences among phases are not exactly $120^o$. Consequently, the approximately balanced three-phase voltages assumption in the lossless and lossy LinDistFlow models deviates from the actual situation of our test cases, leading to higher errors. 

\begin{figure}[t]
    \centering
    \includegraphics[width=3.3in,height=3.3in]{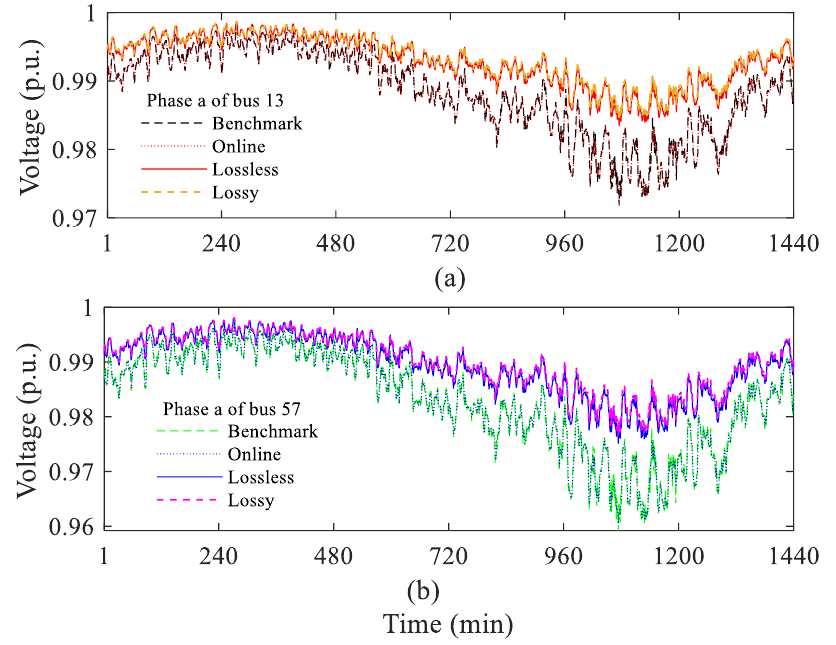}
    \caption{Voltage magnitude comparisons of different models: (a)~Phase a of bus 13; (b)~Phase a of bus 57}
    \label{fig:comparev}
\end{figure}

Taking the voltage magnitudes for phase a of buses 13 and 57 as examples, the voltage magnitude comparisons of different models are shown in Fig.\ref{fig:comparev}. As seen in Fig.\ref{fig:comparev}, the online linearized model has a greater tracking ability and better accuracy compared to the lossless and lossy models across time steps. 
It can be observed that the voltage magnitudes calculated by the online linearized model are very close to the benchmark values, the maximum absolute values of voltage difference between the online model and benchmark for phase a of buses 13 and 57 are only 0.0023 p.u. and 0.0033 p.u., respectively. Taking advantage of the online feedback and closed-loop nature, the online linearized model can capture the time-varying characteristics of unbalanced distribution networks, lending itself to a better approximation to  the exact nonlinear distribution power flow model.
\subsection{Robustness Analysis}
{We test the robustness of the proposed online model against measurement errors, communication failure and update frequencies in this subsection.} 
\begin{figure}[b]
    \centering
    \includegraphics[width=3.3in]{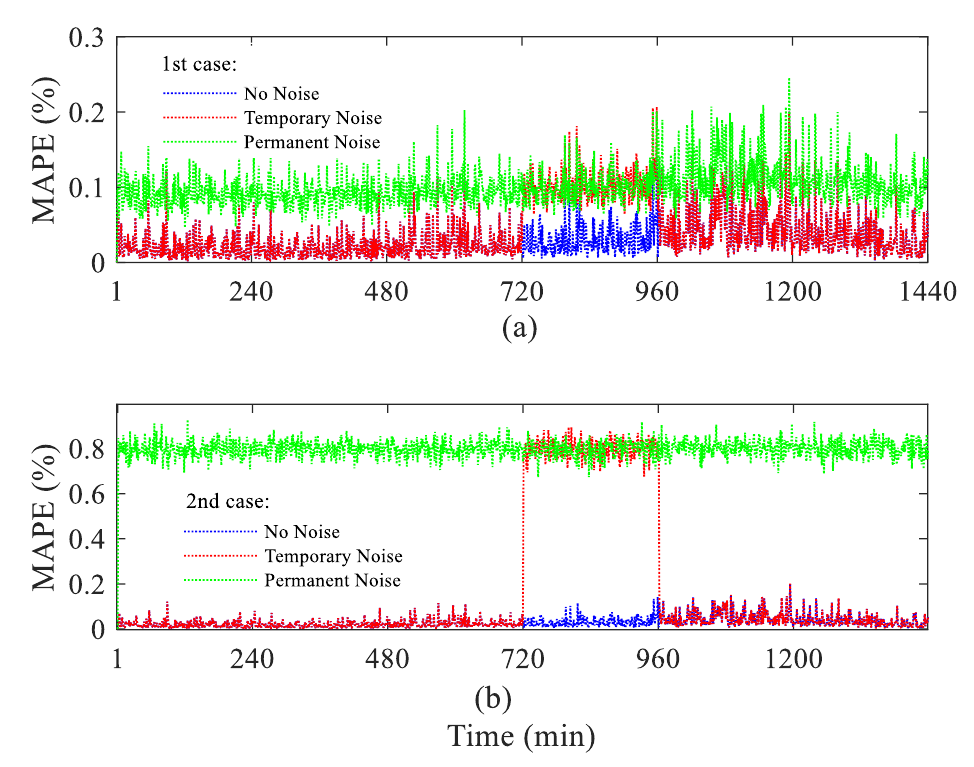}
    \caption{{Robustness against measurement errors: (a) Random noises following a Gaussian distribution $N(0,0.01^2)$ are considered in buses 1, 25, 47, 54, 67, 86, 117, 121; (b) Random noises following a Gaussian distribution $N(0,0.01^2)$ are considered in all the buses across the network}}
    \label{fig:Noise}
\end{figure}

{(1) Robustness against measurement errors. We consider two cases here. In the first case, only the voltage measurements of buses 1, 25, 47, 54, 67, 86, 117, 121 are corrupted by random  noises following a  random Gaussian distribution of $N(0,0.01^2)$. In the second case, each voltage measurement is corrupted by a random  noise following a  random Gaussian distribution of $N(0,0.01^2)$. In the above two 
cases, both the temporary and permanent noises are taken into account, where the temporary noises only exist during 720min-960 min. 
As shown in Fig.{\ref{fig:Noise}}~(a), we can find the degree of oscillations is very small for both temporary and permanent noises when only part of voltage measurements are corrupted with noises. From comparisons between Fig.{\ref{fig:Noise}}~(a) and Fig.{\ref{fig:Noise}}~(b), it can be observed that the degree of oscillations increases as the number of measurements with noises increases. Another point worth mentioning is the temporary noises in both Fig.{\ref{fig:Noise}}~(a) and (b) only lead to deviations during that period. After the period with noises, the online model can return to the state without noises.}

\begin{figure}[t]
    \centering
    \includegraphics[width=3.3in]{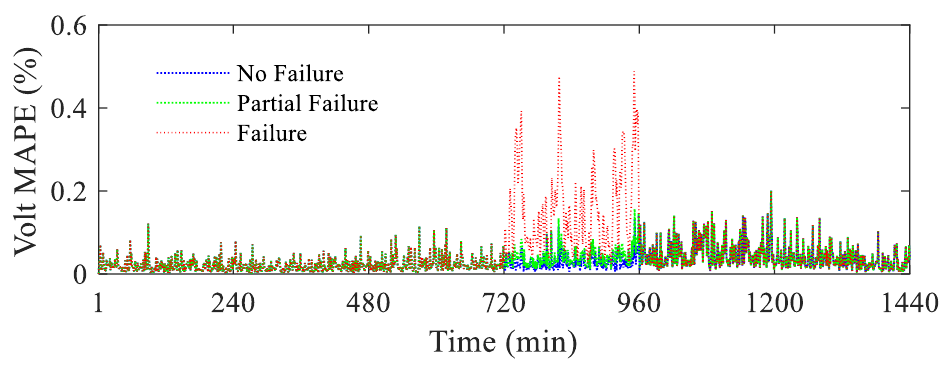}
    \caption{{Robustness against communication failure}}
    \label{fig:Failure}
\end{figure}

{
(2) Robustness against communication failure:
As for communication failure, a ``freeze'' strategy {\cite{JLi}} is used, i.e., the exchanged information remains unchanged until the new value comes. We consider two types of failure: (i) Partial Failure: the communication failure is considered in buses 1, 25, 47, 54, 67, 86, 117, 121 during 720min-960 min; (ii) Failure: the communication failure is considered in  all the buses across the network during 720min-960 min. It can be observed from Fig.{\ref{fig:Failure}} that the degree of oscillations increases as the number of buses with failure increases. And the online model is slightly affected in the Partial Failure case. For both Partial Failure and Failure cases, they can recover to the state without failure after the failure period.
}

\begin{figure}[t]
    \centering
    \includegraphics[width=3.3in]{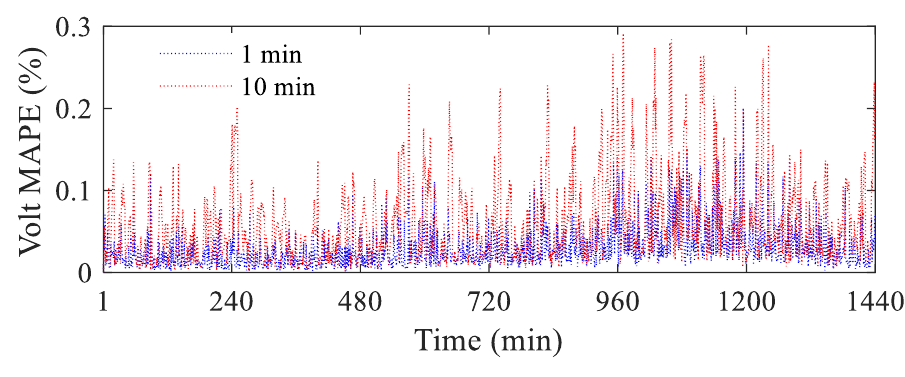}
    \caption{{Robustness against update frequencies}}
    \label{fig:UpdateFrequency}
\end{figure}

{(3) Robustness against update frequencies: We consider two different update frequencies of the online model: (i) the online model is updated every minute, $\Delta{t}=1 \text{min}$; (2) the online model is updated every 10 minutes, $\Delta{t}=10 \text{min}$. As mentioned in Section {\ref{sec:Case}}-A, the time resolution of load data is 1 min. As shown in Fig. {\ref{fig:UpdateFrequency}}, the overall MAPE values of voltage magnitudes for the online model updated every minute are lower than the online model updated every 10 minutes. Since the online model is developed from the FOT expansion that provides a local linear approximation,  a better performance of the online model can be expected with a faster update frequency.
}

\subsection{Delta-Connected Load Analysis}
To further investigate the effectiveness of the online linearized model for delta-connected loads, {the connection type of loads at buses 65 and 76 is changed from wye-connected type to delta-type connected type.} 

\begin{figure}[t]
    \centering
    \includegraphics[width=3.3in]{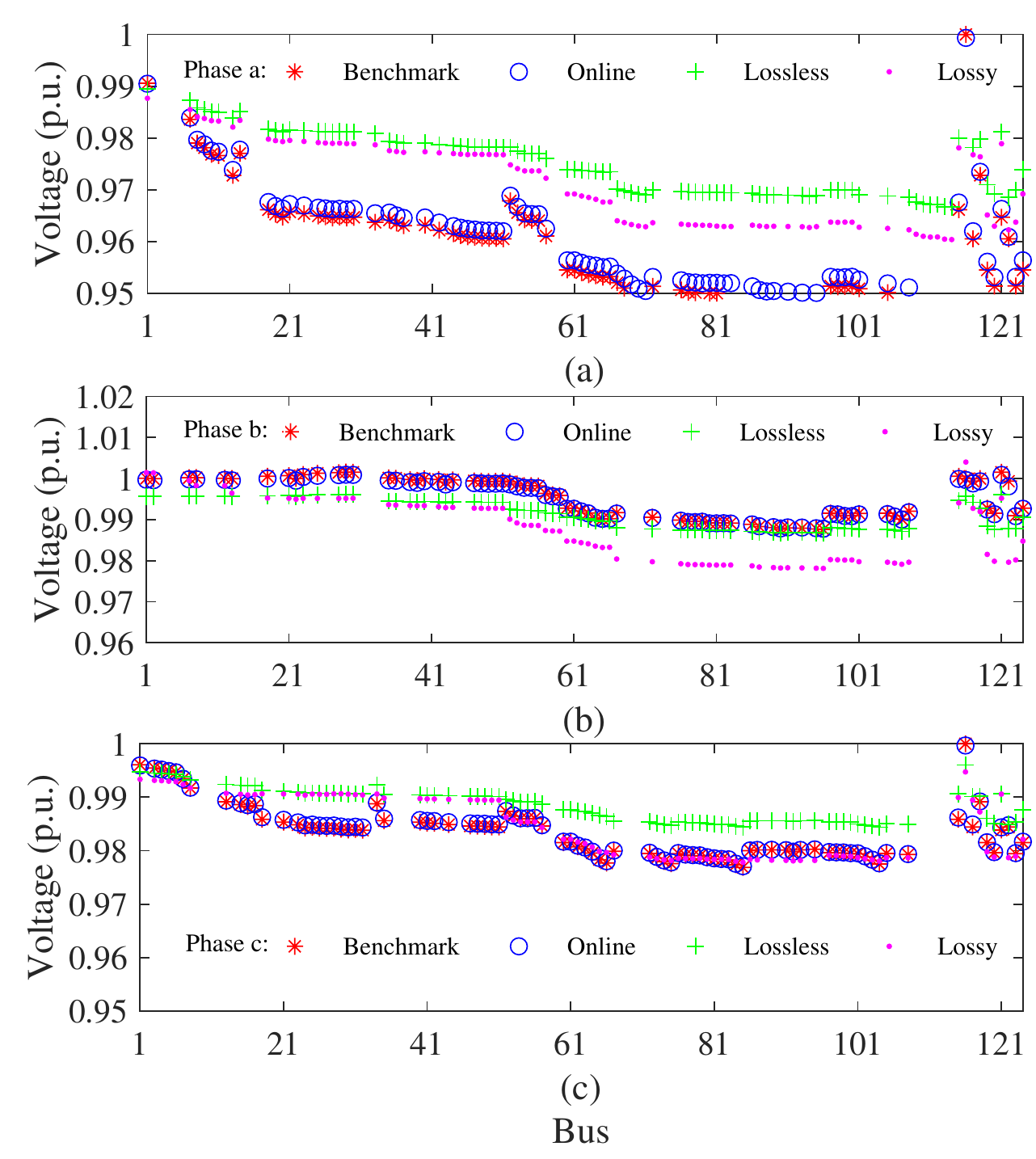}
    \caption{{Phase a, b, and c voltage magnitudes across the buses at the peak load time $t=1074$ min}}
    \label{fig:peakloadtime}
\end{figure}
\begin{figure}[t]
    \centering
    \includegraphics[width=3.3in]{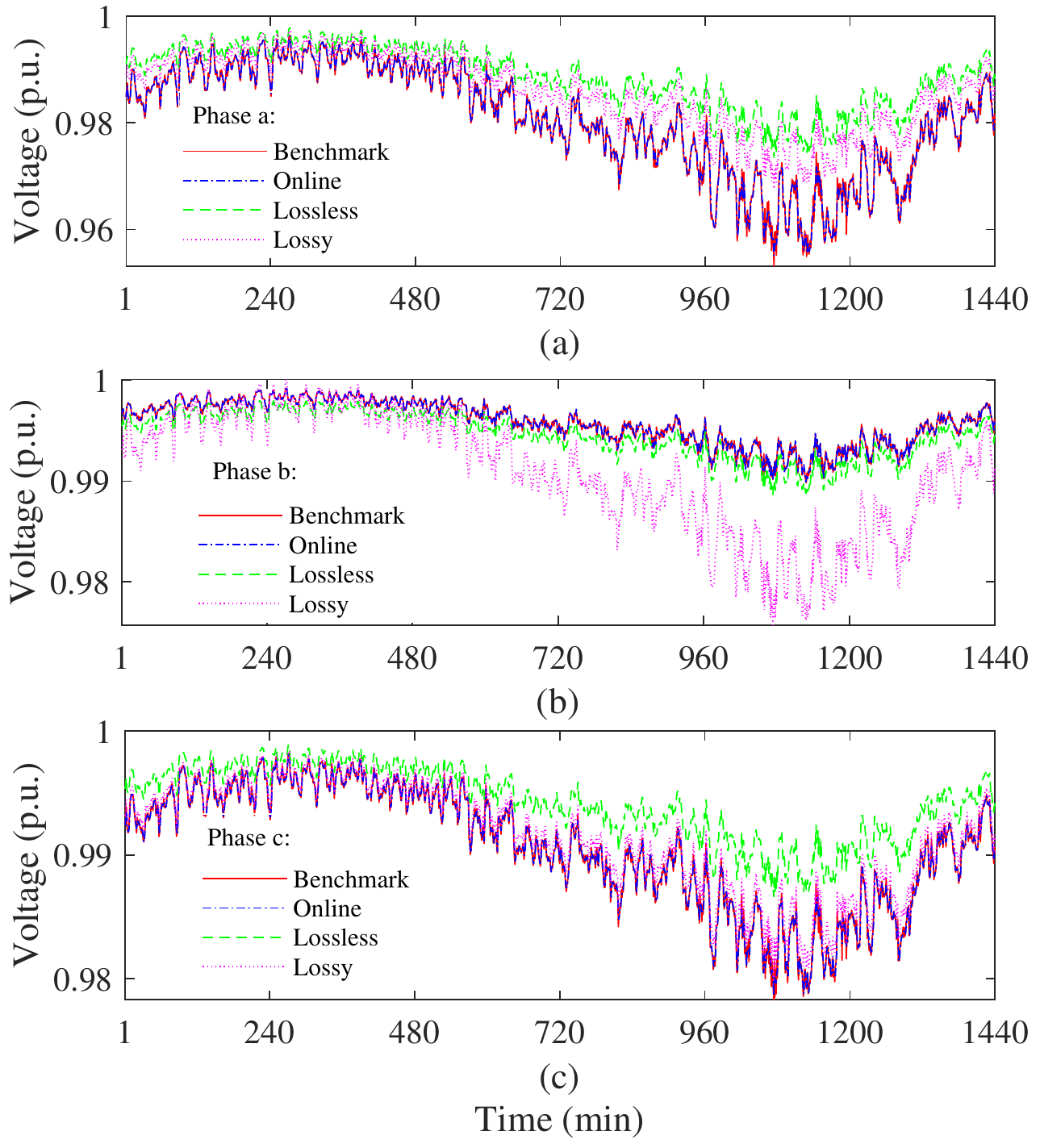}
    \caption{{Voltage magnitudes for bus 65 at each time step: (a)~Phase a; (b)~Phase b; (c)~Phase c}}
    \label{fig:delta65}
\end{figure}

Taking the peak load time $t=1074$ min as an example, Fig.\ref{fig:peakloadtime} shows the phase a, b, and c voltage magnitudes across buses at $t=1074$ min. From Fig.\ref{fig:peakloadtime}, we can find voltage magnitudes across buses calculated by the online linearized model are always closer to the benchmark values for phase a, b, and c compared to the lossless and lossy models. The online linearized model still has a better performance than other models when it comes to unbalanced distribution networks with both wye-connected and delta-connected loads. 

Fig.\ref{fig:delta65} shows the voltage magnitudes for bus 65, associated with the delta-connected load, at each time step. As seen in Fig.\ref{fig:delta65}, the online linearized  model has a great ability to track the benchmark values for bus 65 with the delta-connected load, i.e., the solutions to the exact nonlinear distribution power flow. However, unlike the online linearzied model, the lossless and lossy models cannot track the benchmark values well. It indicates that the online linearized model is also effective for the delta-connected load. The lossless and lossy models are both offline models, which cannot be adjusted with changes in distribution networks. Instead, the transformation matrix $\bm{T}_{\Delta,i}(t)$, given in (\ref{eq:UpdateLoad}), in this proposed online linearized model can easily reflect the time-varying voltage changes for delta-connected loads through online feedback, which results in a better performance for delta-connected loads.
\subsection{Simple Application}
\begin{figure}[t]
    \centering
    \includegraphics[width=3.3in]{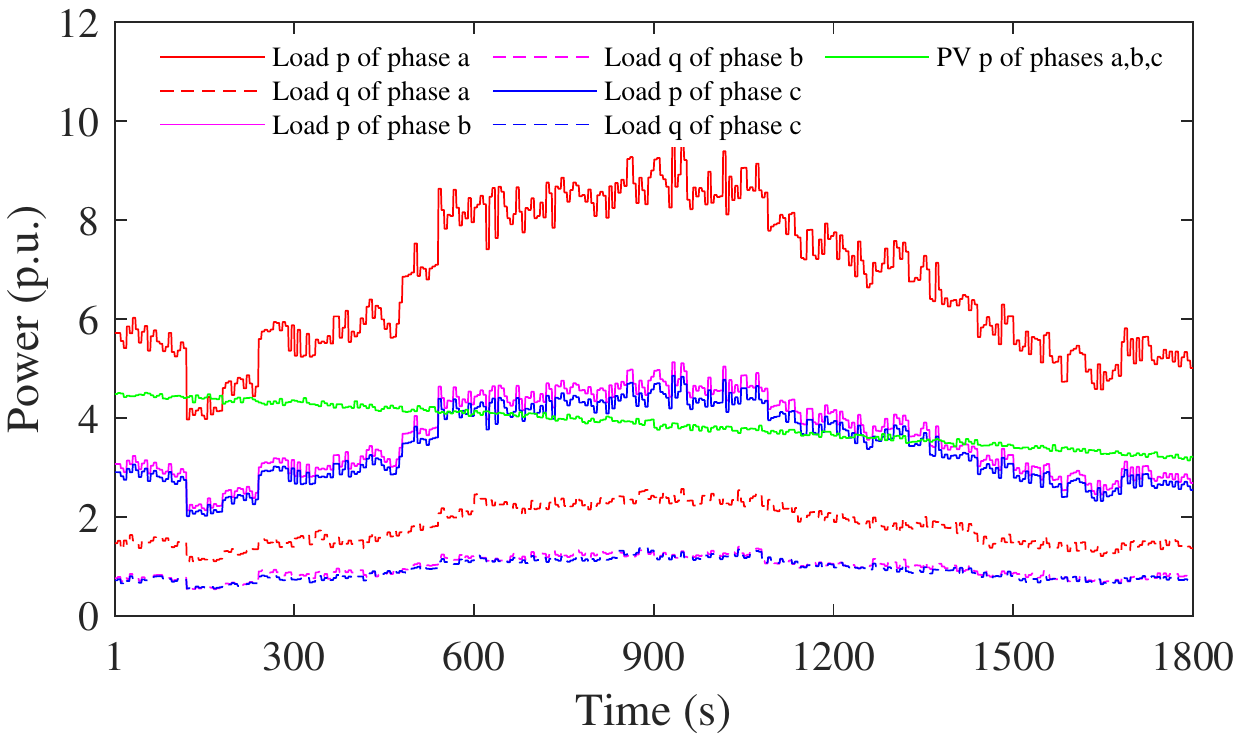}
    \caption{{Aggregate phase load and PV generator profile}}
    \label{fig:OPFload}
\end{figure}

{In this subsection, we apply the proposed model to VVC problem as an example to illustrate its effectiveness. Photovoltaic (PV) generators are installed at buses 13, 29, 48, 50, 56, 60, 66, 79, 83, 95 in the IEEE 123 bus test feeder. With respect to PV generators, it is assumed the real power of PV generators is given and the reactive power is controllable (PV generators can supply or consumer at most 100 kVar reactive power for each phase). We consider a more realistic setting in this simulation, where the time resolution of load and PV generator data is 5 s. The aggregate load and PV generator profile is shown in Fig. {\ref{fig:OPFload}}.  Let $\bm{p}_g$ and $\bm{q}_g$ denote the real and reactive power vectors of PV generators. Here, the VVC program minimizes bus voltage deviations across the network by controlling the reactive power of PV generators, which can be formulated as follows:}
\begin{equation}
    \min f(\bm{v})=\frac{1}{2}||\bm{v}-\bm{1}||_2^2
\end{equation}
{subject to:}
\begin{subequations}\label{eq:optconstr}
\begin{align}
&(\ref{eq:OnlineUpdatePower}a)-(\ref{eq:OnlineUpdatePower}b) \\
\bm{p}&=\bm{p}_{Y}+\bm{K}(t)\bm{p}_{\Delta}^{\Phi\Phi'}-\bm{p}_g\\
\bm{q}&=\bm{q}_{Y}+\bm{K}(t)\bm{q}_{\Delta}^{\Phi\Phi'}-\bm{q}_g\\
&\bm{q}_g^{min}\leq\bm{q}_g\leq\bm{q}_g^{max}
\end{align}
\end{subequations}

{To facilitate the online implementation to solve this problem, we propose the following online strategy to update the reactive power of PV generators at each second $t$, where only one iteration is implemented at each second $t$:

\textbf{[S1]} Collect measurements $(\bm{\hat{V}}(t), \bm{\hat{s}}_{Y}(t), \bm{\hat{s}}_{\Delta}^{\Phi\Phi'}(t))$;

\textbf{[S2]} Update parameters of the proposed online model using ({\ref{eq:UpdateCoefficient}}) and ({\ref{eq:UpdateLoad}}).

\textbf{[S3]} Update the reactive power of PV generators:}
\begin{equation}
    \bm{q}_g(t+1)=\big[\bm{q}_g(t)-\alpha\frac{\partial{\bm{v}}}{\partial{\bm{q}_g}}(\bm{\hat{v}}(t)-\bm{1})\big]_{\bm{q}_g^{min}}^{\bm{q}_g^{max}}
\end{equation}
{where $\alpha$ is the step size, $\frac{\partial{\bm{v}}}{\partial{\bm{q}_g}}$ can be calculated by ({\ref{eq:optconstr}}a)-({\ref{eq:optconstr}}c), $\bm{\hat{v}}(t)$ is the measurement of squared voltage magnitude, calculated by $\bm{\hat{V}}(t)$, $\big[\cdot\big]_{\bm{q}_g^{min}}^{\bm{q}_g^{max}}$ denotes the projection onto $[\bm{q}_g^{min},\bm{q}_g^{max}]$.

\textbf{[S4]} Let $t\leftarrow{t+1}$, and go to \textbf{[S1]}.

The offline OPF calculation is also performed for comparison. There are two main differences between the online OPF and the offline OPF: (i) the constraints ({\ref{eq:optconstr}a})-({\ref{eq:optconstr}c}) of the online power flow model are replaced by the extended LinDistFlow in {\cite{LG}}; (ii) unlike the online implementation, the offline OPF waits until iterations have converged to a solution. With respect to the offline OPF, we assume it is solved every minute by making use of the information obtained at the beginning of every minute.}

\begin{figure}[t]
    \centering
    \includegraphics[width=3.3in]{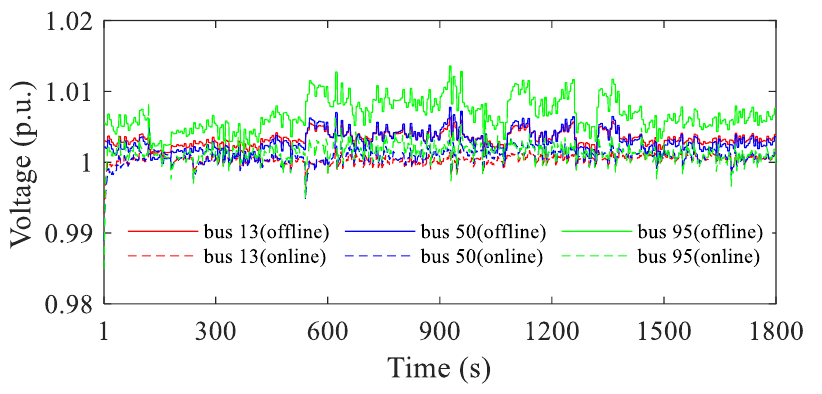}
    \caption{{Phase a voltage magnitudes of buses 13, 50 and 95}}
    \label{fig:OPFV}
\end{figure}

\begin{figure}[t]
    \centering
    \includegraphics[width=3.3in]{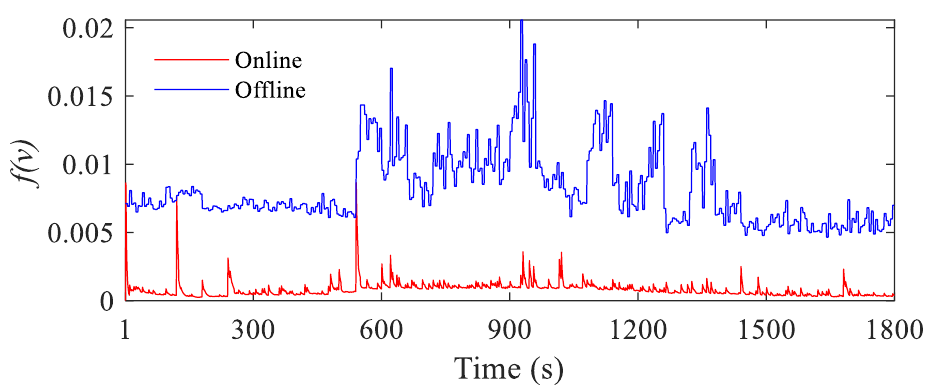}
    \caption{{Comparison of the online and offline OPF for objective function values}}
    \label{fig:OPFComparison}
\end{figure}

{Taking phase a as an example, phase a voltage magnitudes of buses 13, 50 and 95 across time steps are shown in Fig.{\ref{fig:OPFV}}. And comparison of the online and offline OPF for objective function values across time steps is shown in Fig.{\ref{fig:OPFComparison}}. Note that objective function values are obtained by applying the reactive power of PV generators in the actual system where the power flow is solved by the exact nonlinear distribution power flow model. From Fig.{\ref{fig:OPFV}} and Fig.{\ref{fig:OPFComparison}}, we know that phase a voltage magnitudes of buses 13, 50, 95 and objective function values for the online OPF are less than the offline OPF, indicating the better performance of the online OPF. The online implementation can adjust the reactive power of PV generators in real time,  leading to a faster response capability to changes in distribution networks.}



\section{Conclusion}\label{sec:Conclusion}
This paper proposes an online feedback-based linearized power flow model based on the FOT expansion of BFM, for unbalanced distribution networks with both wye-connected and delta-connected loads. By taking advantage of the online feedback, the model parameters can be continuously updated based on the measured voltages and load consumption at the previous time step, to capture  the  time-varying  characteristics of unbalanced  distribution  networks, thus, leading to a compelling performance and tracking ability. Exploiting the connection structure of unbalanced radial distribution  networks, a unified matrix-vector compact form of the proposed linearized power flow mode is also provided for the convenience of online implementation. From the case studies, we have shown that: (i) The proposed online linearized model can track the exact unbalanced distribution power flow very well. (ii)  The proposed online linearized model performs better than the lossless and lossy LinDistFlow models due to its better ability to approximate the nonlinear characteristics. (iii) The proposed online linearized model is applicable to both wye-connected and delta-connected loads. 

{In this work, the model requires the voltage and load measurements at all nodes in the distribution network, which is difficult to be met in reality.  From a practical point of view, some of voltage and load measurements could be replaced by the estimated values with the rapid development of distribution system state estimation (DSSE). In the future, we will explore the combination of DSSE and the proposed online linearized model to reduce the impact of low observability and measurement errors.}
\section*{Appendix A}
\noindent
\textit{Transition between $\bm{s}_{\Delta,i}$ and $\bm{s}^{\Phi\Phi'}_{\Delta,i}$ for open delta-connected loads:}
\medskip

For an open delta-connected load only including the phase-to-phase ab connection, we have:
\begin{subequations}
\begin{align}
    s_{\Delta,i}^{ab}&=(V_i^a-V_i^b)(I_{\Delta,i}^{ab})^*\\
     s_{\Delta,i}^{a}&=V_i^a(I_{\Delta,i}^{ab})^*\\
     s_{\Delta,i}^{b}&=V_i^b(-I_{\Delta,i}^{ab})^*
\end{align}\label{eq:DeltaLoad1}
\end{subequations}
From (\ref{eq:DeltaLoad1}), the open delta-connected load only including the phase-to-phase ab connection can be represented by:
\begin{equation}\label{eq:1}
\begin{split}
    \begin{bmatrix}
    s_{\Delta,i}^a\\
    s_{\Delta,i}^b
    \end{bmatrix}
    =
    \begin{bmatrix}
    \frac{V_i^a}{V_i^a-V_i^b}\\
    -\frac{V_i^b}{V_i^a-V_i^b}
    \end{bmatrix}
    \begin{bmatrix}
    s_{\Delta,i}^{ab}
    \end{bmatrix}
\end{split}
\end{equation}
The open delta-connected load only including the phase-to-phase bc or ca connection can be treated in the same way like the phase-to phase ab connection.

For an open delta-connected load including the phase-to-phase ab and bc connections, we have:
\begin{subequations}
\begin{align}
     s_{\Delta,i}^{ab}&=(V_i^a-V_i^b)(I_{\Delta,i}^{ab})^*\\
    s_{\Delta,i}^{bc}&=(V_i^b-V_i^c)(I_{\Delta,i}^{bc})^*\\
    s_{\Delta,i}^{a}&=V_i^a(I_{\Delta,i}^{ab})^*\\
    s_{\Delta,i}^{b}&=V_i^b(I_{\Delta,i}^{bc}-I_{\Delta,i}^{ab})^*\\
    s_{\Delta,i}^{c}&=V_i^c(-I_{\Delta,i}^{bc})^*  
\end{align}\label{eq:DeltaLoad2}
\end{subequations}
From (\ref{eq:DeltaLoad2}), the open delta-connected load only including the phase-to-phase ab and bc connections can be represented by:
\begin{equation}\label{eq:2}
    \begin{bmatrix}
    s_{\Delta,i}^a\\
    s_{\Delta,i}^b\\
    s_{\Delta,i}^c
    \end{bmatrix}
    =
    \begin{bmatrix}
    \frac{V_i^a}{V_i^a-V_i^b}&0\\
    -\frac{V_i^b}{V_i^a-V_i^b}&\frac{V_i^b}{V_i^b-V_i^c}\\
    0&-\frac{V_i^c}{V_i^b-V_i^c}
    \end{bmatrix}
    \begin{bmatrix}
    s_{\Delta,i}^{ab}\\
    s_{\Delta,i}^{bc}\\
    \end{bmatrix}
\end{equation}
The open delta-connected load only including the phases-to-phase bc and ca connections or the phase-to-phase ca and ab connections can be treated in the same way like phase-to-phase ab and bc connections. 
\section*{Appendix B}

\noindent
\textit{Incidence Matrix Construction for Unbalanced Radial Distribution Networks:}
\medskip

\begin{figure}[htb]
    \centering
    \includegraphics[width=3in]{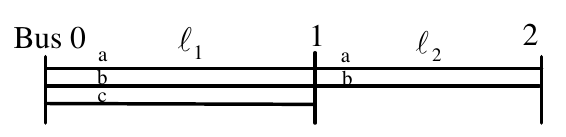}
    \caption{A simple unbalanced radial distribution network}
    \label{fig:Example}
\end{figure}

To illustrate the structure of $\bm{\bar{A}}$, Fig.\ref{fig:Example} shows a simple unbalanced radial distribution network. This simple unbalanced radial distribution network consists of buses 0, 1, 2 and line segments $\ell_1=(0,1)$, $\ell_2=(1,2)$, where buses 0 and 1, $\ell_1$ include phase a, b, c, bus 2 and $\ell_2$ include phase a, b.

For this structure, we can have:
\begin{subequations}
\begin{align}
    \bm{J}(0,\ell_1)&=
    \begin{bmatrix}
    1&0&0\\
    0&1&0\\
    0&0&1
    \end{bmatrix}\\
    \bm{J}(0,\ell_2)&=
    \begin{bmatrix}
    0&0\\
    0&0\\
    0&0
    \end{bmatrix}\\
    \bm{J}(1,\ell_1)&=
    \begin{bmatrix}
    -1&0&0\\
    0&-1&0\\
    0&0&-1
    \end{bmatrix}\\
    \bm{J}(1,\ell_2)&=
    \begin{bmatrix}
    1&0\\
    0&1\\
    0&0
    \end{bmatrix}\\
    \bm{J}(2,\ell_1)&=
    \begin{bmatrix}
    0&0&0\\
    0&0&0
    \end{bmatrix}\\
    \bm{J}(2,\ell_2)&=
    \begin{bmatrix}
    -1&0\\
    0&-1\\
    \end{bmatrix}
\end{align}
\end{subequations}
The incidence matrix $\bm{\bar{A}}$ can be written as:
\begin{equation}\small
\begin{split}
    \bm{\bar{A}}=
    \begin{bmatrix}
    \bm{J}(0,\ell_1)&\bm{J}(0,\ell_2)\\
    \bm{J}(1,\ell_1)&\bm{J}(1,\ell_2)\\
    \bm{J}(2,\ell_1)&\bm{J}(2,\ell_2)
    \end{bmatrix}=
    \begin{bmatrix}
    1&0&0&0&0\\
    0&1&0&0&0\\
    0&0&1&0&0\\
    -1&0&0&1&0\\
    0&-1&0&0&1\\
    0&0&-1&0&0\\
    0&0&0&-1&0\\
    0&0&0&0&-1
    \end{bmatrix}
\end{split}
\end{equation}
with
\begin{equation}\small
\begin{split}
    \bm{{A}}=
    \begin{bmatrix}
    \bm{J}(1,\ell_1)&\bm{J}(1,\ell_2)\\
    \bm{J}(2,\ell_1)&\bm{J}(2,\ell_2)
    \end{bmatrix}=
    \begin{bmatrix}
    -1&0&0&1&0\\
    0&-1&0&0&1\\
    0&0&-1&0&0\\
    0&0&0&-1&0\\
    0&0&0&0&-1
    \end{bmatrix}
\end{split}\label{eq:A}
\end{equation}

\section*{Appendix C}

\noindent
\textit{Proof of Proposition}

Let $\bm{{A}}_a$, $\bm{{A}}_b$, $\bm{{A}}_c$ be the matrices representing the connection structures between each of the buses in $\mathcal{N}$ and each of the line segments in $\mathcal{L}$ in phase a, b, and c radial networks, respectively. For each single-phase radial network, it is a fully connected graph, thus $\bm{A}_a$, $\bm{A}_b$ and $\bm{A}_c$ are all invertible \cite{HZ}-\cite{DBW}. 

Let $\bm{\hat{A}}=\text{blkdiag}(\bm{A}_a,\bm{A}_b,\bm{A}_c)$ be the block diagonal matrix by aligning the matrices $\bm{A}_a,\bm{A}_b,\bm{A}_c$ along its diagonal. We can know $\bm{\hat{A}}$ is invertible, since the following equation holds:
\begin{equation}
    \bm{\hat{A}}\cdot
    \begin{bmatrix}
    \bm{A}_a^{-1}&&\\
    &\bm{A}_b^{-1}&\\
    &&\bm{A}_c^{-1}
    \end{bmatrix}=\bm{I}
\end{equation}
where $\bm{I}$ is the identity matrix. $\bm{{A}}$ can be obtained by $\bm{\hat{A}}$ through elementary row and column operations. Since $\bm{\hat{A}}$ is invertible, $\bm{A}$ is also invertible.

\begin{figure}[htb]
    \centering
    \includegraphics[width=2.3in]{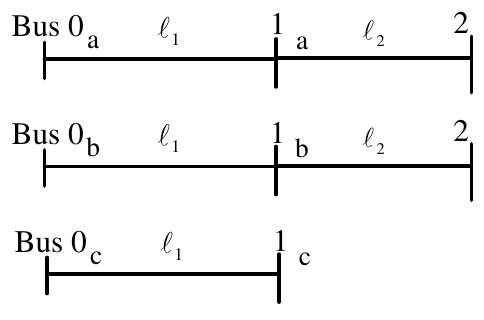}
    \caption{Phase a, b, c radial networks}
    \label{fig:Example2}
\end{figure}

We take Fig.\ref{fig:Example} as an example, the phase a, b, c radial networks of Fig.\ref{fig:Example} are shown in Fig.\ref{fig:Example2}.

In this case, $\bm{A}_a,\bm{A}_b,\bm{A}_c$ can be represented as follows:
\begin{subequations}
\begin{align}
    \bm{A}_a=
    \begin{bmatrix}
    -1&1\\
    0&-1
    \end{bmatrix}\\
    \bm{A}_b=
    \begin{bmatrix}
    -1&1\\
    0&-1
    \end{bmatrix}\\
    \bm{A}_c=
    \begin{bmatrix}
    -1
    \end{bmatrix}
\end{align}
\end{subequations}
Thus, $\bm{\hat{A}}$ can be represented as:
\begin{equation}
\begin{split}
    \bm{\hat{A}}&=\text{blkdiag}(\bm{A}_a,\bm{A}_b,\bm{A}_c)\\
    &=\begin{bmatrix}
    -1&1&0&0&0\\
    0&-1&0&0&0\\
    0&0&-1&1&0\\
    0&0&0&-1&0\\
    0&0&0&0&-1
    \end{bmatrix}
\end{split}
\end{equation} 
The matrix $\bm{A}$, given in (\ref{eq:A}), can be obtained by $\bm{\hat{A}}$ through elementary row and column operations.

\end{document}